\documentclass[10pt,onecolumn,a4paper]{article}
\def\reference{\parskip 0pt\par\noindent\hangindent 0.5 truecm}
\newcommand{\affil}[1]{$^{\rm #1}$}
\usepackage{epsfig}
\date{} 
\title{The History of Astrophysics in Antarctica}
\author{{\it B. T. Indermuehle\affil{A,C}, M. G. Burton\affil{B}, and 
          S. T. Maddison\affil{A}}\\\\
{\small
\affil{A}\,Centre for Astrophysics and Supercomputing, Swinburne University, 
           VIC 3122, Australia}\\
{\small 
\affil{B}\,School of Physics, University of New South Wales, Sydney, 
           NSW 2052, Australia}\\
{\small
\affil{C}\,E-mail: balt@astro.swin.edu.au}
}
\begin{document}
\maketitle
%
%
%
\begin{minipage}{.9\textwidth}
{\bf Abstract}\\
%
We examine the historical development of astrophysical science in Antarctica 
from the early 20th century until today. We find
three temporally overlapping eras, each having a rather distinct
beginning. These are the astrogeological era of meteorite discovery,
the high energy era of particle detectors, and the photon astronomy era
of microwave, sub--mm and infrared telescopes, sidelined by a few niche
experiments at optical wavelengths. The favourable atmospheric and
geophysical conditions are briefly examined, followed by an account of
the major experiments and a summary of their results.

\medskip{\bf Keywords:}
history and philosophy of astronomy -- site testing --
cosmic microwave background -- infrared: general -- sub--millimetre
-- cosmic rays
\medskip
\end{minipage}

%
%

\section{Prehistory}

The first written account of astronomical observations south of the
Antarctic Circle dates back to 1772 (Bayly \& Cook 1782), when the
appointed astronomer and navy officer William Bayly made astrometric
measurements aboard the ships ``Discovery'' and ``Resolution''.  On
their voyage, led by Captain James Cook, they circumnavigated
Antarctica, crossing the Antarctic Circle three times from 1772 to
1775.  The main objective of the observations was to establish
positional accuracy in order to create charts of their discoveries.
For that purpose they were equipped with several different compass
designs, an astronomical quadrant and a Hadley sextant, as well as
accurate chronometers built on the principles of Harrison's ship
chronometer (Andrewes 1996). For stellar and lunar observations, a
Dollond achromatic refractor and a Bird reflector were used. During
the three year long voyage, two lunar eclipses were observed and many
astrometric measurements were made. Their purpose, however, was not to
advance astronomical science but to improve astronomical navigation.

Historical sources are not conclusive on who first sighted the
continent of Antarctica. Among the credible candidates are the Dutch
who sailed the southern sea under Dirck Gerritz (1544--1608).  They
also discovered Tierra del Fuego to be an island and claimed sighting
Antarctica in 1599 as described by Isaac Lemaire in 1622 in the
publication ``Spieghel der Australische Navigatie'' (Mirror of the
Australian Navigation). Other candidates appeared much later (in the
1820's) and include the Russian Thaddeus von Bellinghausen, who saw
the Finibul Ice Shelf and by ``agreement'' is credited with the first
sighting just a few days before the young American sealer 
Nathaniel Palmer and the Briton Edward Bransfield, who sighted Trinity Land 
(Gurney 1998). Although numerous
expeditions were carried out in the 19th century, it appears that none
of them made any science-related astronomical observations. The reason
for this omission is that the scientific disciplines that make
Antarctica attractive as an observational platform today were simply
non-existent or not advanced enough at the time. Additionally,
astronomical equipment that remains functional under the adverse
conditions would not easily have been manufactured and transported to
Antarctica. It is therefore not surprising that the first, and
literally hands-on, astronomical discovery in Antarctica was made in
the unlikely field of astrogeology.

\section{The Early Days}
\label{sec-early}
Sir Douglas Mawson was the leader of the Australasian Antarctic
Expedition, which lasted from 1911 until 1914. Its objective was to
investigate the stretch of Antarctic coast between the then known
boundaries in the west of Terra Nova, which was mapped by Scott's
British Antarctic Expedition in 1910, and Gauss to the east, which was
charted by the German Antarctic Expedition in 1902 (Mawson 1915). A
base camp had been set up at Cape Denison in Commonwealth Bay, Adelie
Land.  It included a transit telescope and hut, whose purpose was for
determining the longitude of Cape Denison through measurement of the
transit times of stars across the local meridian.  However no
astronomy appears to have been done with the telescope.

\begin{figure}[tbh]
\centerline{
\psfig{file=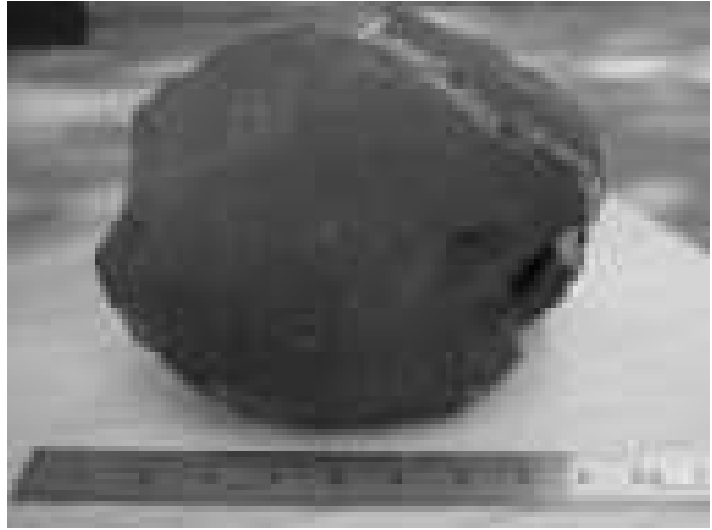,height=7cm}}
\caption{The Adelie Land Meteorite, the first astrophysical find made in 
Antarctica. It was discovered on a sledging party in Douglas Mawson's
Australasian Antarctic Expedition.  The Western Sledging Party, led
by Frank Bickerton, found the meteorite on 5 December 1912, half
buried in the snow, about 18 miles from their base. {\scriptsize
Photograph Michael Burton, October 2002, with acknowledgment to the
Australian Museum, Sydney.}}
\label{Adelie}
\end{figure}

\begin{figure}[tbh]
\centerline{
\psfig{file=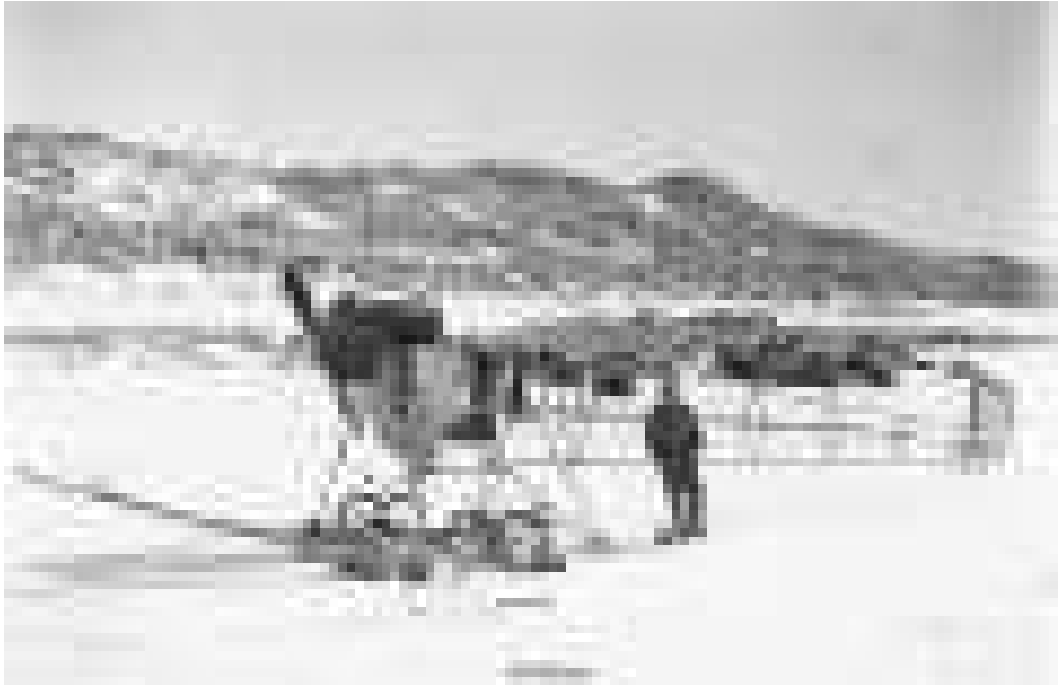,width=14cm}}
\caption{The air tractor used in the 1911--14 Australasian Antarctic 
Expedition, designed for towing of equipment on sleds, with its
engineer Frank Bickerton, discover of the Adelie Land Meteorite.  The
air tractor only functioned for a few hours, however, on Bickerton's
Western Sledging Party, before its pistons seized up. {\scriptsize
Picture credit: Mitchell Library, State Library of New South Wales,
ref.\ ON 144/H475.}}
\label{Airtractor}
\end{figure}

From the base camp, field trips spanning several weeks and extending
significant distances were executed by teams of three. On one of
these excursions, Frank Bickerton led a sledging party with Leslie
Whetter and Alfred Hodgeman. On the third day of the expedition, 18
miles from base and at about 3\,000\,ft elevation, they found a black
object partially buried in the snow (see Figure~\ref{Adelie}). This
chance find, discovered at 12:35 on 5 December 1912, turned out to be
a stony meteorite, about five inches by three inches across (Bayly \&
Stillwell 1923).  It was the first meteorite to be discovered in
Antarctica, and the first astronomical observation of significance on
the new continent\footnote{In an interesting footnote to the story,
the Western Sledging Party also made the first attempt to use
aero-transport in Antarctica with an air tractor -- a light plane on
skis (see Figure~\ref{Airtractor}), with its main wings removed. It
seized up on the second day, with several of its pistons breaking and
smashing the propeller in the process. The trials of mechanising
equipment are one of the great challenges facing Antarctic scientists,
in particular automating experiments for operation at remote sites,
unattended during winter months, as exemplified nearly a century later
with the Automated Astrophysical Site Testing Observatory (the
AASTO) -- see \S\ref{sec-sitetest}.}.  
Bickerton's diary entry for the day gives
a detailed description of the appearance of the meteorite, and makes
it clear that they immediately recognised it for what it was: ``\ldots
meteorite \ldots covered with a black scale, internally of a
crystalline structure, most of its surface rounded except in one place
which looks like a fracture, iron is evidently present in
it''\footnote{Quotation supplied by Frank Bickerton's biographer,
Stephen Haddelsey.  Bickerton's diary is held by the Scott Polar
Research Institute in Cambridge, UK.}.

It had been speculated since the Amundsen and Scott expeditions to the
South Pole in 1911 that the dry and high altitude climate on the
Antarctic plateau might be of advantage to astronomical
observations. The US Navy Rear Admiral R. E. Peary, who had led the
first successful expedition to the North Pole in 1909, was aware of
this and in his relentless quest for discovery addressed his ideas in
a letter to the director of Yerkes Observatory, Professor E. B. Frost
(Peary \& Frost 1912), suggesting that continuous observations during
the course of a year could yield valuable results. He even went on to
estimate the weight limit of the largest equipment part he would
consider transportable, as well as the special construction measures
required on site to enable instruments to be set up on solid ice. 
In 1897 the Yerkes Observatory built a 40 inch refractor telescope,
the worlds largest refractor telescope ever built; it can be imagined 
that Peary sought to contact the Yerkes Observatory thinking they may 
have the research capabilities and interest to pursue astronomy in 
Antarctica. Frost,
however, did not appear to be very enthusiastic about this idea. His
response to Peary's suggestions was sent a mere three weeks later and
clearly stated that the kind of observations that could be made from
South Pole, and which he deemed useful, would not be possible for the
lack of (mainly) precise timing instruments and he seemed generally
reluctant to agree on the pursuit of an observatory at the South
Pole\footnote{There is an interesting footnote to this story. When the
United States established the Center for Astrophysical Research in
Antarctica (CARA), nearly 80 years later, the then director of Yerkes
Observatory, Professor Doyal Harper, became the first director of
CARA.}.

Almost half a century later, the next significant astronomical
discovery was made by Russian geologists operating out of the Lazarev
Station (called Novolazarevskaya today), with a number of meteorites 
found and collected from the Lazarev region in 1961
(Tolstikov 1961; Ravick \& Revnov 1965). The astrogeological
importance of Antarctica, however, did not become evident until 1969,
after a Japanese group of geologists established the first formal
meteorite search programmes based on geological and glaciological
evidence. They successfully retrieved many different kinds of
meteorites including enstatite chondrites, hypersthene achondrites, type III
carbonaceous chondrites and bronzite chondrite in the Yamato region
(Nagata 1975). It was improbable that this accumulation of different
meteorite types in the same location happened by chance, thus driving
the development of theories explaining how meteorites falling in
Antarctica could be transported by the moving ice sheets so as to be
accumulated in a few particular locations --- ablation zones in blue
ice fields --- where they could be easily found (Yoshida et al.\  1971; 
Shima \& Shima 1973).

\section{Antarctica as an Observatory Platform}
\label{sec-platform}
\subsection{Cosmic Rays and Mawson Station}
\label{sec-mawsoncr}
Modern astrophysics was not practiced in Antarctica until the
1950s. Previous astronomical observations undertaken were in a
geophysical context, such as Earth magnetic field observations and
auroral measurements. The first astronomical research programme was
conducted by Australian scientists and developed from cosmic ray
experiments carried out by the Physics department of the University of
Melbourne, initiated by Professor Leslie Martin (Law 2000).  In 1947,
ANARE (Australian National Antarctic Research Expeditions) was
established.  Martin requested that three of his group's experiments
be included in the first ANARE expedition, two to go to the
sub-Antarctic Heard and Macquarie Islands (politically these are in
Australia, not Antarctica, being part of the states of Western
Australia and Tasmania, respectively), and the third to voyage to
Antarctica on the expedition ship Wyatt Earp.  In 1947 and 1948
stations were constructed on Heard and Macquarie Islands and ion
chambers installed there.  The Wyatt Earp continued to Antarctica and
measurements were made using the third ion chamber by Philip Law,
sometimes in the most difficult of circumstances given the frequent
storms of the Southern Ocean.  While not actually taking place on the
continent, this was the first experiment explicitly designed for
astrophysical purposes to be conducted in Antarctica.

In 1949 the cosmic ray equipment was returned from Heard and Macquarie
Island for overhaul, but in 1950 Martin decided to end cosmic ray
research in Melbourne.  The programme moved to the University of
Tasmania, under the direction of Geoff Fenton. Mawson Station,
Australia's first Antarctic station, was established in 1954 and a
cosmic ray observatory built in 1955.  That same year Nod Parsons, a
member of Fenton's group, travelled to Mawson on the Kista Dan and
installed Geiger counters (Parsons 1957).  Three tonnes of lead
absorber were required and packed into 120 boxes on the tiny 1\,200
tonne vessel! These were the first astrophysical experiments to
actually take place on the continent, and the heralded the start of the
Mawson cosmic ray programme which still continues today.

\subsection{The International Geophysical Year}
In 1952, following a suggestion by the US National Academy of Science
member Lloyd Berkner, the International Council of Scientific Unions
(ICSU) proposed to execute a comprehensive series of global
geophysical activities to span the period from July 1957 to December
1958. This was to be called the ``International Geophysical Year''
(Opdishaw 1957), and it was modelled after the International
Polar Years of 1882--1883 and 1932--1933. The intention was to allow
scientists from all over the world to take part in a series of
coordinated observations of various geophysical phenomena. Initially,
46 member states participated and 67 countries had become involved by
the end. In view of the high geomagnetic latitude, the United States 
proposed a cosmic ray detector be located at
McMurdo Station. The US National Science Foundation asked Dr.\ Martin
A. Pomerantz, a leading cosmic ray physicist at the
time, whether he was interested in setting up such a device in
Antarctica (Pomerantz 2000). He accepted and brought with him the great 
experience collected while working on high altitude balloon experiments in the
late 1940s.

A second cosmic ray detector was installed at the South Pole in 1964,
which also provided an opportunity to evaluate the Pole for
other kinds of astronomical observations.  During operations in
1964, a semi-quantitative analysis of stellar and solar observation
conditions were made using a small 3.5~inch aperture telescope. A
report issued in 1970 by the US National Academy of Science and
written by Arne A. Wyller, who was at the time Professor at the Bartol
Research Foundation, concluded that the
seeing would offer excellent conditions for optical astronomy at the
Pole, and suggested the Pole should be considered further as an
astronomical site.  Estimates of the precipitable
water vapour content in the air above the South Pole also indicated that
conditions were very favourable for infrared and millimetre-wave
observations (Mullan, Pomerantz \& Stanev 1989). Nevertheless, no
larger effort was started to take advantage of these conditions for
another decade.

\subsection{The South Pole}
In 1979 the first optical research programme was performed at South
Pole. Eric Fossat and Gerard Grec (both of the Observatoire de Nice),
and Martin Pomerantz coupled a sodium vapour cell to a small telescope
and obtained an unbroken run of over 120 hours of observations
measuring solar oscillations. These data allowed about 80 harmonics of
solar eigenmodes to be discovered (Grec, Fossat \& Pomerantz 1980,
1983; Fossat et al.\ 1987), with periods ranging from about 3 to 8
minutes. In the 1981--82 summer an array detector was added to provide
increased angular resolution for the solar observations, allowing
features as small as ten seconds of arc to be resolved (Pomerantz,
Wyller \& Kusoffsky 1981; Pomerantz, Harvey \& Duvall 1982; Harvey
1989).  The latitude-dependent measurement of the various eigenmodes
of solar oscillation provided evidence that the structure of the
convection zone inside the Sun is different near the equator than it
is at higher latitudes.

In the 1980s, the first projects were envisaged with an initial goal
of observing the galactic infrared emission. Measurements of water
vapour content in the atmosphere made with a site-testing meter in the
summer of 1974 (Westphal 1974) had shown that it was lower than at
Mauna Kea, the world's premier infrared observatory. The first
experiment to attempt to take advantage of this characteristic was a
US--France collaboration that took place in 1984--85 (Pomerantz 1986),
involving Richard Gispert, Jean-Michel Lemarre, Francois Pajot and
Jean Loup Pajot working with Pomerantz. They used a 45\,cm
sub-millimetre telescope named EMILIE (Emission Millimetrique), which
had been designed to work on the 3.6\,m Canada-France-Hawaii Telescope
on Mauna Kea, Hawaii. It was by far the most ambitious and
logistically difficult astronomical programme then undertaken in
Antarctica, requiring the transport of liquid helium to Antarctica all
the way from the USA\@.  The telescope was scanned across the Galactic
plane with a $0.5^{\circ}$ beam, detecting the emission from the
Galactic centre at 900$\mu$m (Pajot et al.\ 1989), and measuring the
dust emission at four wavelengths (460, 720, 850 \& 920$\mu$m) in
several star-forming complexes in the southern Galactic plane,
so allowing the temperature and infrared luminosities of these regions to
be estimated.  Using an upgraded version of EMILIE an attempt was made
in 1986--87, by Mark Dragovan, Tony Stark and Robert Wilson of the
ATT/Bell Laboratories, to measure the cosmic microwave background
anisotropy.  This was to be the first in what was to become an
increasingly sophisticated series of experiments which produced, a
decade later, a series of landmark results concerning the angular
scales and frequency of the anisotropy (see \S\ref{subsec-micro}).

In 1988, the US National Science Foundation awarded a grant for the South 
Pole Astrophysics
Research Center which resulted in a conference held at the University
of Delaware in 1989. The conference proceedings were published (Mullan
et al.\ 1989) and ultimately led to the formation of CARA, the Center
for Astrophysical Research in Antarctica, in 1991 (Harper 1994). CARA 
consisted of a consortium of US universities, and was run from Yerkes 
Observatory in Wisconsin (part of the University of Chicago) under director 
Doyal Harper.

A year round observatory at the South Pole was established by CARA,
with telescopes planned to operate in the infrared, sub-millimetre and
microwave bands. A special ``Dark Sector'', where anthropogenic
interference was to be kept minimal, was set aside 1\,km away from the
Pole for the astronomical experiments.  These have been centred around the
MAPO Building (Martin A Pomerantz Observatory; see Figure~\ref{MAPO}),
named in honour of Pomerantz's many pioneering contributions to the
development of astrophysics in Antarctica. In addition to supporting
several telescopes, MAPO also contains a fully-equipped workshop,
which has proved to be invaluable in maintaining the observatory.  A
programme was also begun to evaluate the transparency, darkness, water
vapour content and stability of the Antarctic sky from infrared to
millimetre wavelengths, for comparison to astronomical sites at
temperate latitudes.

\subsection{Developments within Australia}
The formation of CARA sparked interest in Antarctic astronomy by other
countries, particularly Australia.  Martin Pomerantz first stimulated
discussion by addressing a meeting of the Astronomical Society of
Australia in Hobart in 1986 about the activities then being conducted
at the South Pole (Pomerantz 1986).  This led to Peter Gillingham of
the Anglo Australian Observatory making a presentation in June 1989 to
a meeting at the Academy of Sciences in Canberra, convened to discuss
future plans for astronomy in Australia, on the opportunity offered by
Antarctica. The desirability of joining the establishment of an
international observatory in Antarctica, possibly at the highest point
of the plateau (Dome A), was then publicly discussed at the July 1990
International Astronomical Union Asia--Pacific Regional Meeting held
at the University of New South Wales in Sydney (Gillingham 1991).
Following a government-sponsored visit to Australia by French
scientists in April 1991, collaboration in Antarctic astronomy was
viewed as a high priority.  A return visit to France the following
year established links with the Universit\'{e} de Nice group and the
idea born to measure micro-thermal turbulence in order to quantify the
seeing.  Gillingham and Jean Vernin (of the Universit\'{e} de Nice)
then visited CARA in Wisconsin to advance plans for making
micro-thermal measurements. A visit to John Bally of the University of
Colorado by Michael Burton in September 1992 further established links
with CARA, and a collaboration was formed to measure the infrared sky
brightness using a recently superseded infrared photometer of the
Anglo Australian Observatory (the IRPS -- see \S\ref{sec-sitetest}).
Jamie Lloyd, who was at the time completing his undergraduate degree
at the University of New South Wales, and Michael Burton took these
two experiments to the South Pole in January 1994, the experiments
having largely been put together by then graduate student Rodney Marks
and his supervisor Michael Ashley at UNSW\@. By the end of that year,
the Joint Australian Centre for Astrophysical Research in Antarctica
(JACARA) was established between the University of New South Wales in
Sydney and the Australian National University in Canberra to
facilitate further cooperation with CARA (Burton et al.\ 1996).

\subsection{International Developments}

At the 21st General Assembly of the International Astronomical Union (IAU), 
held in Buenos Aires in 1991, a working group chaired by Peter Gillingham 
was established to encourage the development of Antarctic astronomy. 
Seventeen papers were presented at the meeting (see Gillingham 1992) and 
a resolution encouraging international collaboration in Antarctic astronomy 
was drafted and adopted by the General Assembly. 
At the 1994 IAU General Assembly in the Hague, a
full-day session was held on the topic, with the chair of the working
group passing to Michael Burton.
A couple of weeks later in Rome a special session on Antarctic
Astronomy was held at the 22nd SCAR meeting (Scientific Committee for
Antarctic Research).  This meeting was organised under the auspices of STAR, 
the Solar Terrestrial and Astrophysical Research working group of SCAR, with
John Storey (also of the University of New South Wales) becoming the
vice-chair with responsibilities for astrophysics within STAR\@.  SCAR
also passed a resolution recognising the scientific value of Antarctic
astronomy and calling for the development of the field.  Antarctic
astronomy meetings have been regular features of IAU and SCAR meetings
ever since. STAR has since been re-organised as the Standing
Scientific Group on Physical Sciences (SSG/PS), with both an expert
group (AAA -- Antarctic Astronomy \& Astrophysics) and an action group
(PASTA -- Plateau Astronomy Site Testing in Antarctica), providing it
with input.

\section{The Advantages of the Antarctic Plateau for Astronomy}
\label{sec-advantage}
At the South Pole, with its relatively high altitude of 2\,835\,m above sea
level and an equivalent pressure altitude of 3\,200\,m due to the
extreme cold, the transparency of several atmospheric windows is
greatly improved. This is largely due to the extremely low levels of
precipitable water vapour in the atmospheric column, which regularly
falls to below 250\,$\mu$m ppt H$_2$O, compared to levels of
$\sim$1\,mm at the best mid-latitude sites (Townes \& Melnick 1990).

The advantages of the Antarctic plateau for astronomy at infrared
wavelengths also stem from its extremely cold temperatures,
stable atmosphere, and the ability to observe objects continuously
throughout the long winter night (Harper 1989; Storey \& Hyland 1993;
Burton et al.\ 1994; Burton 1996; Ashley 1998; Storey 2000). The sky
at the South Pole at thermal infrared wavelengths is darker than at
any mid-latitude site by between one and two orders of magnitude,
and thus dramatically reduces the background noise and minimises the sky
fluctuations (Harper 1989; Burton, Allen \& McGregor 1993; Nguyen et
al.\ 1996; Ashley et al.\ 1996; Phillips et al.\ 1999; Chamberlain et
al.\ 2000).  

In 1991 Peter Gillingham suggested that high sites on the Antarctic
plateau might also provide unprecedented seeing (``super-seeing''),
based on the absence of diurnal temperature cycles in winter and the
slow settling of air from the stratosphere over the plateau
(Gillingham 1993a, 1993b). This would occur above a shallow surface
inversion layer, where the air at the ice surface is cooler than at
the top of the inversion layer by up to $\sim 20^{\circ}$C, typically
only 200--300\,m higher.  Turbulent mixing in this surface inversion
layer has been confirmed as causing the relatively poor surface seeing
($\sim$1.5" in the visual) at the Pole.  However, because there is
little thermal inhomogeneity (other than in the air near the
telescope) and because the air moves relatively slowly, the
isoplanatic angle and coherence time are both greatly increased over
their values at temperate-latitude sites. This provides conditions
which are very favourable for adaptive correction of wavefronts (Mark
2002).  Since there is no jet stream contributing to high-altitude
turbulence at the Pole, the seeing is dominated by the contribution
from this narrow surface layer.  Not only has this significance for
attempts to recover the diffraction limit from the turbulence induced
seeing, but also for conducting astrometric interferometry, reducing
the phase errors due to the proximity of the turbulence cells to the
telescope, compared with the high altitude turbulence encountered at
mid-latitude observatories (Lloyd, Oppenheimer \& Graham 2002).  On
the summits of the Antarctic plateau, where wind speeds are at their
lowest (as this is where katabatic winds originate), even the surface
seeing contribution is expected to be very small at times.
Experiments are now being conducted at the high plateau site of Dome C
to quantify this attribute (Travouillon et al.\ 2003c).

In the cosmic ray spectrum the Earth's magnetic field normally shields
(or at least deflects) a significant amount of the heavier charged
particles travelling through space. At the magnetic Poles the magnetic
field lines enter the surface of the Earth almost vertically, thus
creating a port of entry for charged particles.  Thus Antarctica is a
particularly suitable place to study cosmic rays.

Another significant advantage in Antarctica is found in the vast
amount of transparent ice over the plateau, reaching to a depth of
3--4\,km below the surface.  This can be used to create a neutrino
particle detector of enormous volume, which is needed in order to be
able to record the minuscule number of neutrinos that do in fact
interact with nuclei in the ice, or underlying rock, as they pass
through the Earth.

\section{Site Testing at the South Pole}
\label{sec-sitetest}
The first testing to measure the near-IR sky brightness from
2--5$\mu$m was conducted from 1994--1996 using the Infrared
Photometer Spectrometer or IRPS (Ashley et al.\ 1996; Phillips et
al.\ 1999), originally used as the Anglo-Australian Telescope's
front-line IR instrument in the early 1980s\footnote{When it was used,
among other things, to discover a cluster of hot young stars in the
centre of the Galaxy (Allen, Hyland \& Jones, 1983), and to find
windows in the infrared bands which could be used to peer through the
atmosphere of Venus to see the tops of the highest mountains (Allen \&
Crawford 1984).}. At the same time, the 60\,cm SPIREX telescope was
also used to measure the sky brightness from 1--2.5$\mu$m. From
2.3--5$\mu$m the sky was found to be between 10 and 100 times darker
than at temperate latitude sites.  Microthermal turbulence
measurements were also conducted from a 30\,m high tower (in 1994) and
from balloons (in 1995) to characterise the contribution of the
boundary layer to atmospheric seeing (Marks et al.\ 1996, 1999).

\begin{figure}[tbh]
\centerline{
\psfig{file=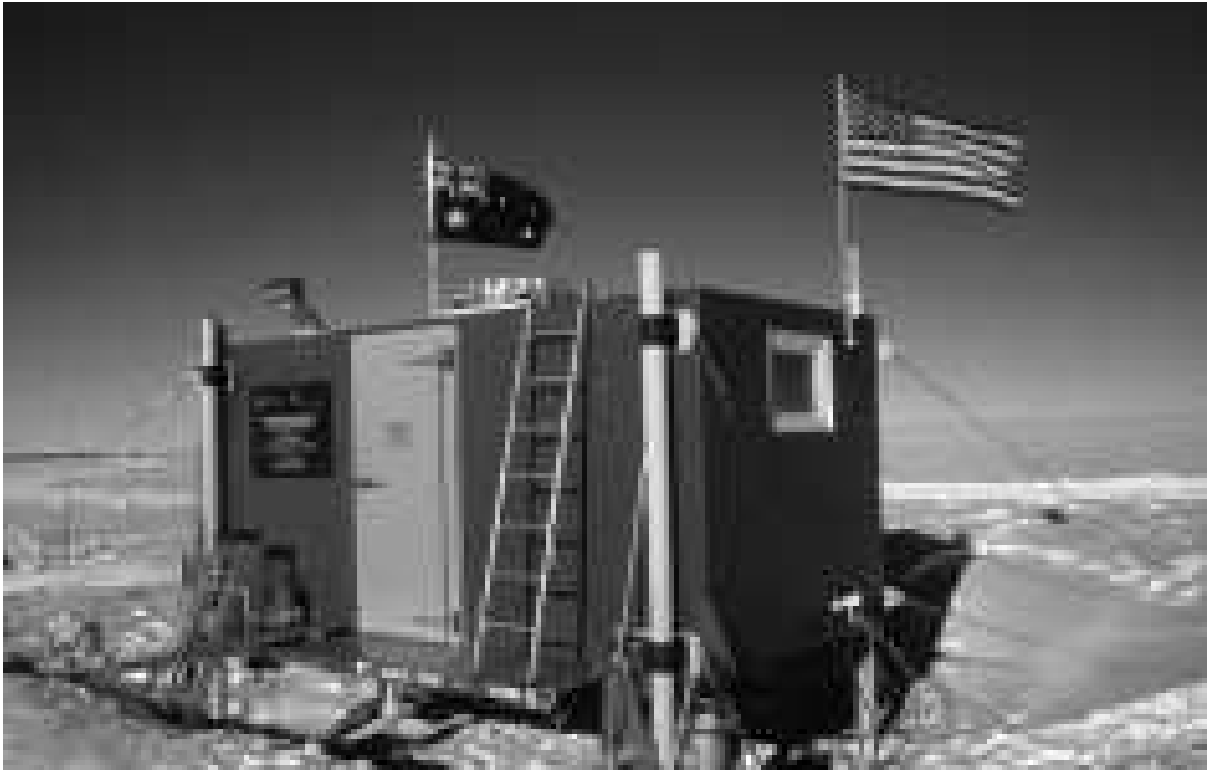,width=16cm}}
\caption{The Automated Astrophysical Site Testing
Observatory (AASTO), installed at the South Pole in 1997. This
self-powered, autonomous laboratory, hosted a suite of site testing
instruments for the purpose of quantifying the performance of the site
for astronomical use. {\scriptsize Image provided by Michael Burton,
January 1999.}}
\label{AASTO}
\end{figure}

In order to better provide site characteristics and to facilitate
further site evaluation, the Automated Astrophysical Site Testing
Observatory (AASTO; see Figure~\ref{AASTO}) was established at the
South Pole in 1997 (Storey et al.\ 1996). This self-powered,
autonomous laboratory, hosting a suite of site testing instruments,
was developed as a practical solution to the challenge of obtaining
data in a harsh remote environment with limited electrical power and
with no human intervention. Experiments were designed and built for the AASTO
to measure the electromagnetic spectrum from the UV to the
sub-millimetre in order to quantitatively assess the site for
astronomical use. The AASTO project began a collaboration between the
University of New South Wales and the Australian National University,
in conjunction with CARA, and led to the formation of JACARA (Burton
et al.\ 1996). It continued to operate at the South Pole until the end
of 2003, when it became the platform for conducting an experiment to
search for extrasolar planetary transits (SPETS -- the South Pole
Exoplanet Transit Search; Caldwell et al.\ 2003).

The AASTO programme had to overcome considerable technical challenges,
not the least being the difficulty of obtaining a reliable power
supply able to operate without interruption or maintenance over the
winter months. The thermo-electric generator (or TEG) used to provide
power is driven via a catalytic oxidisation of propane fuel.  This
catastrophically failed several times during the first three years of
operation, leaking freon which decomposed over the oxidiser,
producing hydrochloric and hydrofluoric acid. The net result was
considerable chemical erosion of the instrument suite (Storey, Ashley
\& Burton 2000). Without some considerable efforts by the UNSW team over
the summer months, plus a series of heroic efforts by successive
winter-over scientists (Paul Sullivan, Michael Masterman and Charlie
Kaminski), the AASTO programme would not have succeeded, and likely the
whole site testing programme would have been put back several years. In
January 2003, an improved version of the laboratory was transported to,
and deployed, at Dome C -- the AASTINO (Antarctic Astrophysical Site
Testing International Observatory; Lawrence et al.\ 2003), where it
provided the first wintering facility at the new Concordia Station,
currently under construction by the French and Italian national
Antarctic programmes.

The AASTO package consisted of a suite of instruments, developed from
1997--2001, including the Near-Infrared Sky Monitor (NISM) and the
Mid-Infrared Sky Monitor (MISM) (Storey et al.\ 1999), which were both
used to measure sky brightness in the respective wavebands (Lawrence
et al.\ 2002; Chamberlain et al.\ 2000); the Antarctic Fibre-Optic
Spectrometer (AFOS), used to measure atmospheric transmission from the
UV to the far-red (240--800\,nm; Boccas et al.\ 1998, Dempsey et al.\
2003b); a Sonic Radar (SODAR), used to measure the level of turbulence
in the surface inversion layer (Travouillon et al.\ 2003a); the
Antarctic Differential Image Motion Monitor (ADIMM), used to determine
the astronomical seeing by measurements of stars (Travouillon et al.\
2003b); and the Sub-millimetre Tipper (SUMMIT), used to measure the
350\,$\mu$m sky brightness (Calisse et al.\ 2004).

In addition to the astronomical site testing data, the AASTO also
collected weather data such as temperature, wind speed, wind direction
and atmospheric pressure.  The AASTO instruments measured the sky
brightness from 1.25--14\,$\mu$m and recorded the incidence of clear
skies suitable for astronomical observations.  Around 50\% of all
`nights' at the Pole were found to be of excellent quality. In the
near-infrared, the lowest background levels were measured in a window
between 2.3\,$\mu$m and 2.45\,$\mu$m, where it attained levels of less
than 70\,$\mu$Jy/arcsec$^2$, values comparable to measurements made
from high altitude balloons and two orders of magnitude less than at
mid-latitude sites. Residual emission in this window is thought to
come from airglow at altitudes above 38\,km and, interestingly, no
correlation was observed between the sky brightness and auroral
activity. At longer wavelengths (from 3--30\,$\mu$m), while the
background reduction was found to be only one order of magnitude, the
advantages of Antarctica were found to be especially significant. In
these bands infrared arrays rapidly saturate due to the high thermal
backgrounds, atmospheric windows are only partially transparent, and
sky fluctuation noise is high, limiting the stability of
measurements. All these attributes were found to be improved at the
Pole. For wavelengths shorter than 2.3\,$\mu$m, however, the gains are
fairly modest because the sky brightness is dominated by OH airglow
rather than thermal emission, and this varies little around the
world. Since the quality of the observing conditions at wavelengths
longer than 3\,$\mu$m depends principally on temperature, it did not
come as a surprise that good observing conditions were found to extend
well beyond the boundaries of the polar night.

\section{High Energy Physics}
\label{sec-highenergy}
\subsection{Mawson Station}
As described in \S\ref{sec-mawsoncr}, in 1955, two muon telescopes
were built and shipped to Mawson Station on the Antarctic coast and
installed by Nod Parsons (Parsons 2000). The design for each telescope
was based on three trays of Geiger counters, each 1\,m$^2$ in size,
placed inside a specially constructed building -- the cosmic ray
hut. The relative ease of operation made them suitable for the first
experiments to be located in Antarctica and cosmic ray detectors are
now installed at many Antarctic research stations. These first
telescopes still required a significant amount of maintenance,
however, and a faulty paper recorder meant that measurements of the
great solar flare event of 23 February 1956 were compromised. In
December 1956, a 12-counter neutron monitor was sent to Mawson to be a
part of the Australian Neutron Monitor Network (McCracken 2000). It
has been a great contributor to measurements of many solar flare
events since 1957, most notably the event of 4 May 1960, where it
provided experimental verification of the spiral nature of the Solar
System magnetic field long before direct measurements of the field by
satellites (McCracken 1962). The monitors have also had a crucial role
to play in the calibration of space-based measurements.

In 1968, observations commenced at Mawson with two high zenith angle
muon telescopes, pointing north and south at 76$^{\circ}$ (Jacklyn
2000).  One significant result was finding a latitude-dependent
sidereal semi-diurnal variation (Jacklyn \& Cooke 1971). A new cosmic
ray observatory was constructed in 1971, designed largely by Attila
Vrana who was an electronics engineer with the Australian Antarctic
Division (AAD).  A vertical shaft was driven through the granite rock
to a depth of 40\,mwe (metres of water equivalent) to filter out the
lower energy particles. Two vaults were constructed at the bottom, one
of which would harbour an underground muon telescope. On top of this
was the building containing the high zenith angle telescopes and the
neutron telescope. The detectors used, however, remained as Geiger
counters until their replacement in 1982 by proportional
counters. These telescopes found an intriguing solar modulation at the
time of a very large cosmic ray decrease in July 1982 (Jacklyn, Duldig
\& Pomerantz 1987).  These variations became known as isotropic
intensity waves, but since there have been no clear recurrences of the
original phenomenon they remain something of a mystery.  Further
upgrading continued during the 1980's, under the direction of Marc
Duldig, to fully automate the system and transfer the data by
satellite back to the AAD headquarters in Hobart, Tasmania.  To date,
the muon detector at Mawson remains the only such southern hemisphere
telescope at Polar latitudes, along with one in Tasmania (Duldig
2002). A full historical account of the Australian cosmic ray
programme can also be found in this paper.

\subsection{South Pole}
The GASP telescope (Gamma Astronomy at the South Pole; Morse \& Gaidos
1989) was installed at the Pole from 1994-97 by the University of
Wisconsin-Madison.  Six metre-sized
mirrors were used to look for the Cherenkov light from cosmic rays
which are generated by gamma ray interactions in the upper atmosphere.
The long, dark night, combined with the constant zenith angle that an
astronomical source would have, were the primary reasons for
installing the telescope at the Pole, following the building of a
similar facility at Haleakala Observatory in Hawaii.  However, the
facility was not successful with its objective of finding celestial
sources of gamma rays, as no clear detections of discrete sources were
ever made.

Cosmic rays with energies of $\sim 10^{14}$\,eV are measured by large
collecting area air shower arrays. At these high energies the flux of
events is too low to measure with short duration balloon flights, and
the collecting area too small for expensive satellite experiments.
The constant zenith angle of any astronomical sources at the South
Pole greatly simplifies the analysis of the air shower data, and large
collecting area experiments can readily be built there. Two such
instruments have been built at the South Pole Station to measure
cosmic rays -- SPASE and SPASE--2\@.

SPASE, the South Pole Air Shower Array, was built just 200\,m away
from the South Pole and near to the Geodesic Dome. It was established
in 1987 and ran continuously for 10 years. SPASE was built under the
direction of Alan Watson (University of Leeds) and Martin A. Pomerantz
(the same Pomerantz who built the first cosmic ray detectors at
McMurdo in 1961)\footnote{As an interesting footnote, he went on to
own a Nissan and later a Ford car dealership near Huntsville, Alabama,
for about ten years.}  and was a joint effort between the Bartol
Research Institute of the University of Delaware and the cosmic ray
group of the University of Leeds (Smith et al.\ 1989). Sixteen
scintillation detectors were spread over a collecting area of
6\,200\,m$^2$. The experiment specifically aimed to find cosmic sources
of gamma ray emission.  The trajectories of these photons are not
altered by the galactic magnetic field, unlike the case for charged
cosmic ray particles.  However no such sources were ever found (van
Stekelenborg et al.\ 1993), and the particle events detected had an
isotropic distribution across the sky.

Construction started in 1994 on an enhanced array called SPASE--2 (Dickinson
et al.\ 2000) in the ``Dark Sector'' of South Pole Station. 
SPASE--2 was built on top of the Antarctic Muon And
Neutrino Detector Array (AMANDA, see below), which was simultaneously
also under construction. The new air shower facility began operation
in 1996, with a larger array than SPASE, containing 120 scintillator
modules spread over an area of roughly 16\,000\,m$^2$. An additional 9
air-Cherenkov telescopes were placed around them, a separate
experiment given the name of VULCAN\@. SPASE--2 is sensitive to cosmic
rays with primary energies from $10^{14}$ to $3 \times 10^{16}$\,eV\@.
It was also built with the express purpose of working in concurrence
with the AMANDA neutrino detector to measure the electron component
of air showers while AMANDA measured their muon component.

\begin{figure}[tbh]
\centerline{
\psfig{file=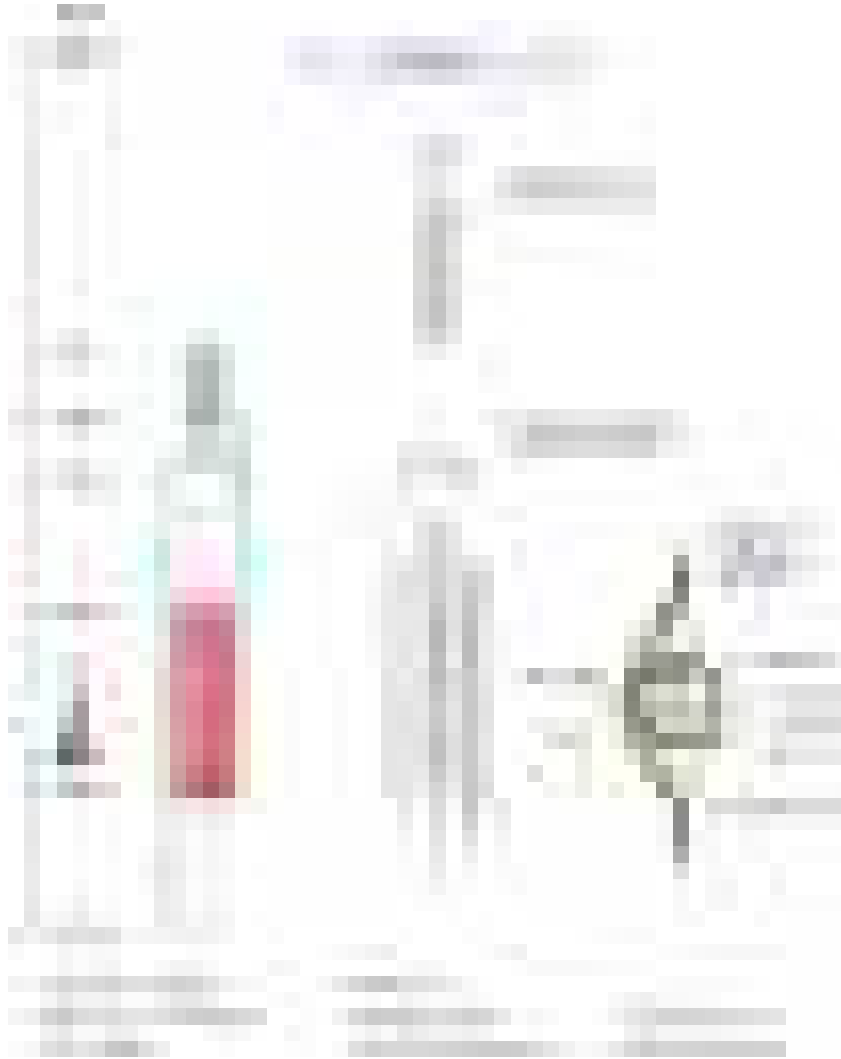,height=14cm}}
\caption{Schematic diagram of AMANDA, the Antarctic Muon and Neutrino
Detector Array, at the South Pole.  AMANDA contains 677 widely spaced
optical modules, connected together by strings and placed into
water-filled holes that were drilled into the ice, before the water
re-freezed.  Their depths range from several hundred metres to 3\,km
deep.  {\scriptsize Image from the AMANDA consortium, courtesy of
Christian Spiering.}}
\label{AMANDA}
\end{figure}

The AMANDA project has become the largest single scientific programme
at the South Pole, with collaborating scientists from 20 institutions
in the USA, Sweden, Germany, Belgium, the UK and Venezuela (Andres et
al.\ 2000).  AMANDA contains widely spaced photomultiplier tubes
(PMTs; see Figure~\ref{AMANDA}), connected together by strings and
placed into water-filled holes drilled into the ice.  The holes range
from several hundred metres to 3\,km in depth and the water in them
rapidly re-freezes, trapping the PMTs within the ice. High energy
neutrinos coming up through the Earth will {\it very} occasionally
interact with nuclei in the ice or rock and so create a muon, which in
turn emits Cherenkov radiation when passing through the ice. By
measuring the arrival times of the light pulses at the PMTs, the
origin of the neutrinos and their arrival rate can be determined. In
addition, they may possibly be brought into coherence with events
measured by SPASE\@. The detectors in AMANDA point downwards, to look
for neutrinos that enter the Earth in the northern hemisphere and pass
right through it, before encountering a nucleus in the ice.  They
point down in order to shield the detector from the vastly greater
count rate generated by downward passing cosmic rays.  The first four
strings of PMTs were deployed over the 1993/94 summer period
(AMANDA--A), but at the 800--1\,000\,m depth they were placed at, the
ice was found to contain too many air bubbles to allow the muon tracks
to be followed. Since then the array has been expanded twice, with 19
strings placed at depths from 1\,500--2\,000\,m (AMANDA--II), and
containing 677 optical modules.  The great transparency of the ice at
these depths, where the absorption length reaches to $\sim 100$\,m,
and the absence of biological contaminants, makes the Antarctic ice
particularly suitable for such an experiment. So far over 1\,500
neutrino events have been detected, isotropically distributed over the
northern sky (Andres et al.\ 2001, Wiebusch et al.\ 2002).

\subsection{Neutrino Experiments Under Development} 
No individual sources of neutrino emission have yet been detected,
however.  Nevertheless, the sensitivity thresholds of AMANDA are much
as expected, with about three atmospheric neutrinos detected per day.
It is anticipated that the next generation neutrino telescope,
IceCube, will be able to detect individual cosmic sources of neutrino
emission, such as active galactic nuclei and gamma ray bursters, thus
opening up a completely new field of observational astronomy (Halzen
2004).  IceCube will use a cubic kilometre of ice as its collecting
volume, opening up unexplored bands for astronomy including the PeV
(10$^{15}$\,eV) energy region.  At such high energies the Universe is
opaque to $\gamma$-rays originating from beyond the edge of our own
galaxy, whereas cosmic rays of this energy do not carry directional
information because of their deflection by magnetic fields.

Two experiments aimed at detecting ultra high energy neutrinos (i.e.\
$> 10^{20}$\,eV) through their interaction with nuclei in the
Antarctic icecap are also under development. At such extreme energies the
only particles capable of reaching the Earth from cosmological
distances are neutrinos. Electrons, scattered into the particle
cascade when the neutrinos interact with the ice, emit a pulse of
Cherenkov radiation, peaking in the radio at frequencies of a few
hundred MHz.  A prototype experiment called the Radio Ice Cherenkov Experiment,
or RICE, operated at the South Pole over the 1995/96 and 1996/97
summers (Allen et al.\ 1998) using two of the AMANDA bore holes. It is
being developed further to work with the full AMANDA array.  Another
experiment under development is ANITA, the ANtarctic Impulsive
Transient Antenna, scheduled to be deployed from a long duration
balloon flying at 40\,km altitude. Ice is transparent to radio waves
of frequencies $\sim 1$\,GHz, so that ANITA will be sensitive to
neutrino-generated radio pulses occurring from a large part of the
Antarctic ice sheet, equivalent to having an effective telescope
collecting area of 1 million square kilometres (Barwick et al.\ 2003)!
ANITA is planned to be launched from McMurdo in 2005.


\section{Photon Astronomy}
\label{sec-photon}
We divide photon astronomy into four bands, defined by four atmospheric
windows that are open, for our purposes of describing the various
astrophysical experiments undertaken in Antarctica.  These are the
optical, infrared, sub-millimetre and microwave windows.

\subsection{Optical}
\label{subsec-optical}
\begin{figure}[tbh]
\centerline{
\psfig{file=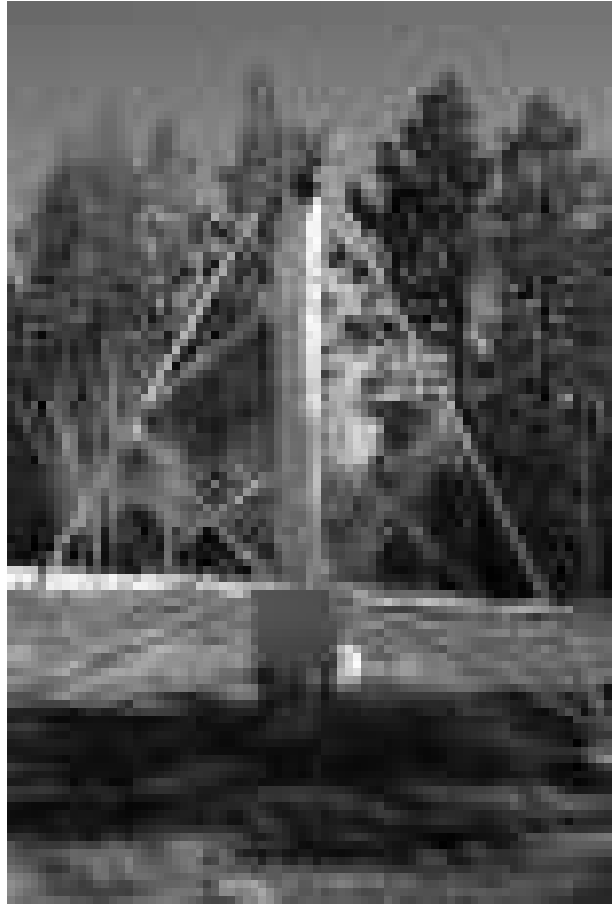,height=12cm}}
\caption{The SPOT telescope, a 2 inch telescope which was 
the first optical telescope to be operated through the winter at the
South Pole.  It is shown as it is currently installed, at the Sunspot
Solar Observatory, New Mexico, USA. {\scriptsize Photograph Balthasar
Indermuehle, November 2002.}}
\label{SPOTPiccy}
\end{figure}

Optical astronomy has mainly been employed for site testing and
evaluation in Antarctica, with limited astronomical applications. The
South Pole Optical Telescope (SPOT, a 2 inch periscope style
telescope; see Figure~\ref{SPOTPiccy}) programme was operated by the
University of Florida from 1984 to 1988 (Chen et al.\
1987). Measurements were taken during the austral summer to evaluate
the visual seeing conditions. Long period measurements were also made
in winter, obtaining up to one week of continuous data of the light
curves of variable stars, including the Wolf-Rayet star
$\gamma^2$\,Velorum. However, the data was affected by clouds, and the
useful observing time did not exceed that of mid-latitude
observatories (Taylor 1990).

Optical telescopes have also been used to measure the visual seeing at
night at the Pole, through the use of DIMMs (differential image motion
monitors; Bally et al.\ 1996; Dopita, Wood \& Hovey 1996). These
monitor light from a star through several adjacent apertures in order
to measure the stability of the atmosphere. Two such experiments were
the HDIMM on the SPIREX telescope (Loewenstein et al.\ 1998) and the
ADIMM used with the AASTO (Travouillon et al.\ 2003b).  The ice-level
seeing at the Pole was found to be relatively poor, of order $1.5''$,
but since it is nearly all produced in a narrow surface inversion
layer, the prospects for adaptive optic correction are extremely
promising (Marks 2002; see \S\ref{sec-advantage}).

Argentinean astronomers examined site conditions at General Belgrano
II Station, located at 79$^\circ$\,S on the coast. They used a
Celestron CG11 telescope and measured an average $3.8''$ viewing,
which is not good enough for serious observations. The extinction
across the optical bands was also found to be high. The U band, however, may
potentially be interesting for observation of auroral activity and
effects associated with the ozone hole (Mosconi et al.\ 1990).

\subsection{Infrared}
\label{subsec-infrared}
\begin{figure}[tbh]
\centerline{
\psfig{file=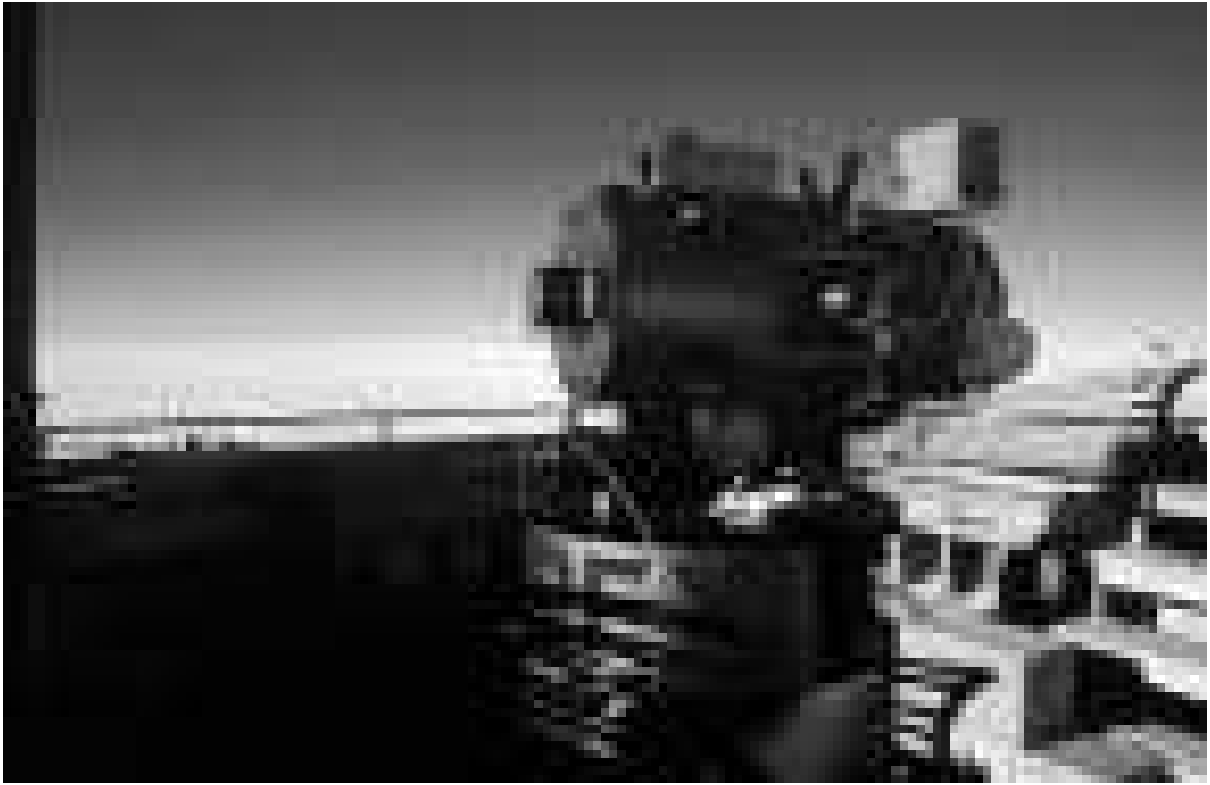,width=16cm}}
\caption{SPIREX, a 60\,cm infrared
telescope, which operated at the South Pole and was sensitive to
radiation from 1--5\,$\mu$m. It was installed just in time to witness
the collision of Comet Shoemaker-Levy 9 with Jupiter in July 1994, the
only telescope in the world with the opportunity to continuously
observe the week-long series of impacts. SPIREX was decommissioned in
2000. {\scriptsize Image provided by Michael Burton, January 1999.}}
\label{SPIREX}
\end{figure}

The South Pole Infrared Explorer (SPIREX; see Figure~\ref{SPIREX}),
was a 60\,cm telescope observing from 1--5\,$\mu$m at the South Pole.
It operated with two different instruments -- the GRism IMager (GRIM, 
from 1993 until 1997; Hereld 1994) and the Abu camera (in 1998 and 1999; 
Fowler et al.\ 1998). SPIREX was installed just in time to witness the 
collision of Comet Shoemaker-Levy 9 with Jupiter in July 1994, the only
telescope in the world with the opportunity to continuously observe
the week-long series of impacts.  A total of 16 impacts were
recorded (Severson 2000).

SPIREX's principle scientific work (see Rathborne \& Burton (2004) for a
fuller description) was for imaging PAHs (polycyclic aromatic
hydrocarbons) at 3.3\,$\mu$m, tracers of photodissociation regions
(the surface of molecular clouds excited by far--UV radiation from
young stars). PAHs were imaged in NGC\,6334 (Burton et al.\ 2000) and
in the Carina nebula (Brooks et al.\ 2000; Rathborne et al.\
2002). Extensive shells of PAH emission were found, wrapped around
embedded protostellar objects.

SPIREX also was used to measure infrared excesses at 3.5\,$\mu$m from
hot dust in disks around pre-main sequence stars. Infrared excesses
from disks were searched for in the nearby low mass star forming
clouds of Chamaeleon I (Kenyon \& Gomez 2001) and $\eta$
Chaemaeleontis (Lyo et al.\ 2003).  At 3.5\,$\mu$m the signature from
disk emission was found to be much more readily detectable than in the
normally used 2\,$\mu$m band.  Despite its modest size, the SPIREX
telescope obtained the then deepest image at 3.5\,$\mu$m from any
telescope, achieving a detection threshold of 18.2 magnitudes per
square arcsecond in a 10 hour observation of the 30 Doradus region of
the Large Magellanic Cloud (Rathborne \& Burton 2004).

\subsection{Sub-millimetre}
\label{subsec-submm}
Heavy elements are created in nuclear processes in the interiors of
stars and returned to the interstellar medium through winds and
supernova.  Carbon is the most abundant of these heavy elements,
emitting at 370\,$\mu$m and 610\,$\mu$m in the submillimetre, and
plays a significant role in many astrophysical processes. Atomic
carbon observations are difficult from Earth because water vapour
makes the atmosphere opaque to much of the sub-millimetre band. At the
South Pole these windows are open and make it possible to observe
neutral carbon as a matter of routine.

\begin{figure}[tbh]
\centerline{
\psfig{file=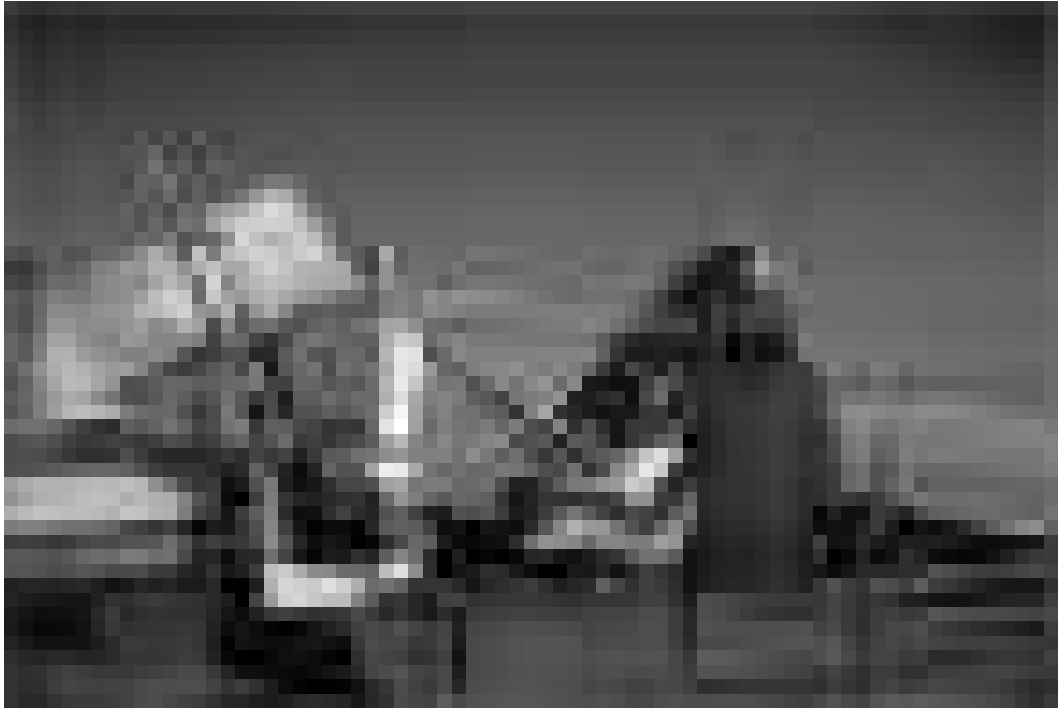,width=14cm}}
\caption{Rich Chamberlin in front of the 1.7\,m diameter AST/RO dish. The 
main mirror focuses sub--mm wavelength radio waves and directs them,
by a series of mirrors, into a warm room below the telescope. AST/RO
has been the most scientifically productive telescope in Antarctica,
with over 50 refereed publications resulting from its
operation. {\scriptsize Photograph copyright Smithsonian Astrophysical
Observatory.}}
\label{ASTRO}
\end{figure}

AST/RO, the Antarctic Sub-millimetre Telescope and Remote Observatory
(see Figure~\ref{ASTRO}), is a 1.7-metre telescope built for this
purpose, with the primary aim of measuring the dominant cooling lines
from dense interstellar gas, where stars are forming (Lane \& Stark
1997; Stark et al.\ 2001). Observation of these lines allows the
temperature and density of the gas to be determined.  AST/RO has been
in almost continuous operation since January 1995, making it the
longest serving astronomical telescope on the Antarctic plateau to
date. It is a general-purpose facility, used both for astronomy and
aeronomy at wavelengths between 200--2000\,$\mu$m, and has been
equipped with several detector systems (and given an eclectic set of
names -- Major Dobbin, Wanda, FLaMR, PoleSTAR, TREND and SPIFI, as well
as having a Fourier transform spectrometer for aeronomy studies; see
Table 1). The optics are offset to produce high beam efficiency and to
avoid inadvertent reflections and resonances. The primary reflector is
made of carbon fibre with a surface accuracy of about 9\,$\mu$m. It
has Coud\'{e} and Nasmyth focii. Most of the spectroscopy performed
has used heterodyne receivers mounted on an optical table in the warm
coud\'{e} room. AST/RO operates continuously throughout the winter
period. The science carried out has concentrated on measurement of
emission lines of atomic carbon and carbon monoxide from
photodissociation regions, in both the Milky Way and the Large
Magellanic Cloud, which is well placed for observation from the South
Pole.  AST/RO also contributed to the South Pole site evaluation
programme, by measuring characteristics of the sub-millimetre sky
emission.

The major accomplishments of AST/RO to date include the first
detection of [CI] emission in the Magellanic Clouds (Stark et al.\
1997); a survey of [CI] and CO emission from the HII region/molecular
cloud complexes of Carina (Zhang et al.\ 2001) and NGC 6334 (Yan et
al.\ 2004); a survey of the [CI] emission from high galactic latitude
molecular clouds (Ingalls et al.\ 2000) and from the inner few degrees
of the Galaxy, to determine the CI/CO ratio as a function of
metallicity, and to compare with the LMC/SMC and the solar
neighbourhood (Ojha et al.\ 2001; Martin et al.\ 2004).


The all-year operation, as well as the high degree of automation,
requiring only one winter-over astronomer to execute the observations,
and the warm environment for the instruments, have all contributed to
the success of AST/RO\@. It has been the most productive telescope
built in Antarctica.

The Submillimetre Polarimeter for Antarctic Remote Observing (SPARO)
was a 9-pixel 450\,$\mu$m polarimetric imager developed by scientists
at Northwestern University in Chicago for use with the Viper
telescope, where it was installed from 2001--2003. SPARO has been used
to determine the direction of interstellar magnetic fields through the
polarisation of dust emission in the sub--mm, for instance mapping the
magnetic field through the inner several hundred parsecs of the Milky
Way.  As a result a large-scale toroidal magnetic field was detected
over the region of the Galactic centre (Novak et al.\ 2003).

\subsection{Microwave}
\label{subsec-micro}
The Cosmic Microwave Background Radiation (CMBR) spectrum resembles
that of a blackbody at a temperature of 2.7\,K. It carries a wealth of
information about the origin and evolution of the Universe in its
signal, allowing predictions of the standard Big Bang model to be
rigourously tested. The angular power spectrum of the CMBR contains
information on the structure that existed at decoupling (the time when
radiation and matter separated). Anisotropy in the matter distribution
at this time can also be detected in the CMBR as minuscule variations
in the angular distribution from a smooth 2.7\,K blackbody
spectrum. The amplitude and spatial distribution of the anisotropy are
directly related to the conditions in the early Universe which
ultimately gave rise to the formation of structure, such as the
super-clustering of galaxies.  Due to the extreme stability of the
atmospheric microwave emission, Antarctica has proved to be the
pre-eminent Earth-based site for CMBR measurement, and has hosted a
series of increasingly sophisticated experiments as a result.

In the late 1980s, a ground-based experiment was carried out by an
Italian research team in Terra Nova Bay to measure
anisotropies in the CMBR at an angular scale of 1.3$^\circ$. A 1\,m
diameter flux collector with a He$^3$ cooled bolometric detector,
sensitive to radiation with a wavelength between 1.86 and 2.34\,mm,
was used. Observations were carried out during the Antarctic summer
period (dall'Oglio et al.\ 1988).

At about the same time the South Pole was also chosen as a site for
several CMBR investigations.  After some trial experiments the first
major campaign was conducted during the 1988-89 summer.  It took place
in what was dubbed ``Cucumber land'' by the South Pole support staff
in view of how the experiment appeared to them!  They were set up in
three Jamesway tents (hence the appearance of a ``cucumber'' -- a
long, cylindrical green tent) about 1 mile grid south east from the
Geodesic Dome, and involved two research groups -- from Princeton led
by Jeff Peterson and from the University of California, Santa Barbara,
led by Phil Lubin.  Several experiments were involved.  A one square
metre Bell Labs offset horn antenna was used with a single-pixel 1\,mm
wavelength bolometer (called ``Miss Piggy''), borrowed from
the University of Chicago airborne astronomy programme.  UC Santa
Barbara provided a balloon gondola and Princeton trialled the ``White
Dish'' experiment.

It took a few years before the CMBR experiments began obtaining
results.  The first significant result came using the White Dish
(Tucker et al.\ 1993), from measurements taken between 1991 and 1993
which provided tight limits on the anisotropy at higher angular scales
than had been probed with the COBE satellite (ie.\ $\Delta T < 63
\mu$K for $\theta \sim 0.15^{\circ}$).  White Dish comprised of a
1.4\,m telescope with a single-mode waveguide bolometer operating at
90\,GHz. With the formation of CARA, CMBR observations at the Pole
became incorporated into the COBRA programme (COsmic microwave
Background Radiation Anisotropy).  COBRA's first experiment was the
Python telescope, built in 1992. This 0.75\,m instrument used three
detector systems to map the CMBR at 38, 42 and 90\,GHz,
respectively. The CMBR structure analysis provided some of the best
confirmed data of the anisotropies on degree angular scales at the
time (Coble et al.\ 1999). By including many repetitions and
variations of the same measurement, Python showed that the results
were reproducible and not instrumental in origin. To help discriminate
between the various models of structure formation, it was essential to
increase both the areal sky coverage and the range of angular scales
probed by Python's observations. The Viper telescope extended studies
to arcminute scales and higher frequencies, thus superseding Python.

\begin{figure}[tbh]
\centerline{
\psfig{file=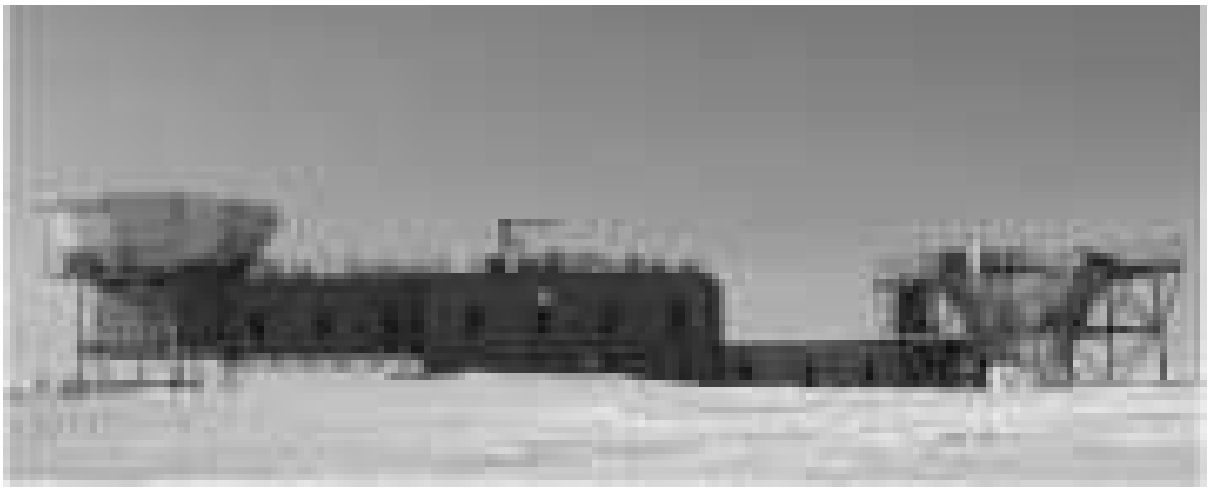,width=16cm}}
\caption{The Martin A Pomerantz Observatory (MAPO). The 60\,cm SPIREX 
infrared telescope was originally hosted on the tower on the the
left. In 2000 the DASI CMBR telescope was installed in its place. To
the right is the Viper telescope, a 2.1\,m off-axis telescope which
hosted the ACBAR and SPARO instruments, used for CMBR studies and for
mapping polarisation at sub-mm wavelengths, respectively. {\scriptsize
Image provided by the Office of Polar Programs, National Science
Foundation.}}
\label{MAPO}
\end{figure}

Viper, a 2.1 metre off-axis telescope, was installed at the South Pole
in January 1998\@ (Peterson et al.\ 2000; see Figure~\ref{MAPO}). The
optical design included an electrically driven mirror which allowed
the beam to be rapidly swept across the sky by several degrees,
without significant beam distortion or large variations in telescope
emission and ground emission pick-up. This large sweep ability allowed
Viper to be used to make observations that extended across several
degrees, considerably improving on the data obtained with
Python. Among the instruments deployed on Viper were SPARO (see the
sub--mm section, above) and ACBAR, the Arcminute Cosmology Bolometer
Array Receiver, a 16 element bolometer array cooled to just 250\,mK\@
that was deployed in 2000 (Runyan et al.\ 2003).  ACBAR operates at
three frequencies centred on 150, 219 and 279\,GHz, extending the
spectral range over which the CMBR anisotropy has been measured. As
with all the CMBR experiments, multiple frequencies are necessary in
order to be able to separate the CMBR signal from foreground sources
of contamination, produced mostly by dust, synchrotron and free-free
radiation. ACBAR was designed specifically to make higher angular
resolution measurements of the CMBR anisotropy, from $l = 200$ to 3000
(Kuo et al.\ 2003). It also aimed to measure the Sunyaev-Zeldovich
Effect (SZE) in galactic clusters. The SZE is produced by hot (several
million K) gas in galaxy clusters, which Compton scatters the CMB
radiation, modifying its blackbody spectral distribution.  Data from
ACBAR was also combined with data from the Cosmic Background Imager
(CBI) and added to that collected by the Wilkinson Microwave
Anisotropy Probe (WMAP) to provide the current best estimates on
cosmological parameters such as the Hubble constant, the age of the
Universe, and the contributions of ordinary matter, cold dark matter
and dark energy to the overall composition of the Universe (Spergel et
al.\ 2003).

DASI, the Degree Angular Scale Interferometer, 
is a 13 element interferometric array designed to
extend the angular coverage of the CMBR to values from $140 < l <
910$ (Leitch et al.\ 2002; see Figure~\ref{MAPO}). 
DASI specifically complemented the Viper telescope, especially
the mm and sub--mm capabilities provided by ACBAR\@. DASI made the
first detection of polarisation in the CMBR (Kovac et al.\ 2002), a
result not only making the front cover of the Christmas issue of
Nature in 2003, but also the front page of the New York Times. These
results provided strong support for the underlying theoretical
framework that explained the generation of CMBR anisotropy, and lent
great confidence to the values of cosmological parameters derived from
anisotropy measurements.

\begin{figure}[tbh]
\centerline{
\psfig{file=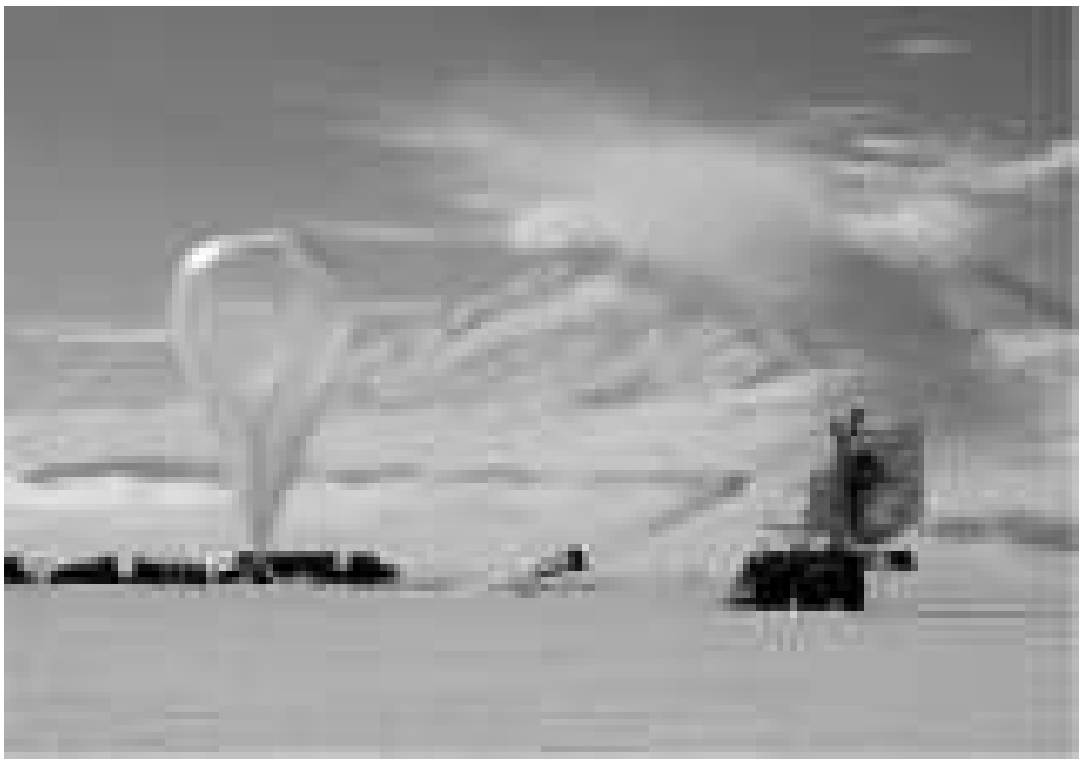,height=10cm}}
\caption{The BOOMERanG (Balloon Observations Of Millimetre Extragalactic
Radiation and Geomagnetics) experiment, being launched from the US
coastal station McMurdo in December 1998.  BOOMERanG carried a 1.2\,m
microwave telescope to an altitude of 38\,km by a balloon, and was the
work of a US/Italian team. {\scriptsize Image provided by Office of
Polar Programs, National Science Foundation, 1998.}}
\label{BOOMERANG}
\end{figure}

The BOOMERanG (Balloon Observations Of Millimetre Extragalactic
Radiation and Geomagnetics; see Figure~\ref{BOOMERANG}) experiment used a
1.2\,m microwave telescope carried to an altitude of 38\,km by
balloon launched from the US coastal station McMurdo by a
US/Italian team. It combined the high sensitivity and broad frequency
coverage, pioneered by an earlier generation of balloon-borne
experiments, with 10 days of integration time obtained on a
long-duration balloon flight. Such long periods are possible because a
nearly circular pattern of east-to-west jet-stream winds establishes
itself in the Antarctic stratosphere for periods of a few weeks at the
end of summer. The circulation is generated by the long-lived
high-pressure system over the continent caused by the constant solar
heating of the stratosphere in summer. This allows, in principle, the
launching and recovery of a balloon from roughly the same geographic
location and permits a flight path that is almost entirely over
land. Constant temperature due to the permanent day conditions during
austral summer also permit the balloon to maintain an essentially
constant altitude, as well as minimising any temperature fluctuations
in the observing platform. The data retrieved from BOOMERanG (de
Bernardis et al.\ 2000) were inconsistent with the then current
cosmological models based on topological defects but were consistent
with a subset of cold dark matter models. They provided the best
evidence at the time that the geometry of the Universe was indeed flat
(i.e.\ Euclidean). As with the DASI result two years later, these
measurements also made the front cover of Nature, in April 2000.

\section{The Future}
\label{sec-future}
The South Pole is nearing completion of a major upgrade of its
infrastructure, with the construction of a new Station to replace the
Geodesic Dome installed in the 1970's. Two major new astronomical
facilities are under development to exploit this enhanced capability
for supporting science, the 10\,m South Pole Telescope (SPT; Carlstrom
2003), to be used to measure the SZ--effect towards galaxy clusters in
the sub--mm, and IceCube (Halzen 2004), a 1\,km$^3$ collector volume
neutrino telescope, capable of imaging the high energy neutrino sky
for the first time.

\begin{figure}[tbh]
\centerline{
\psfig{file=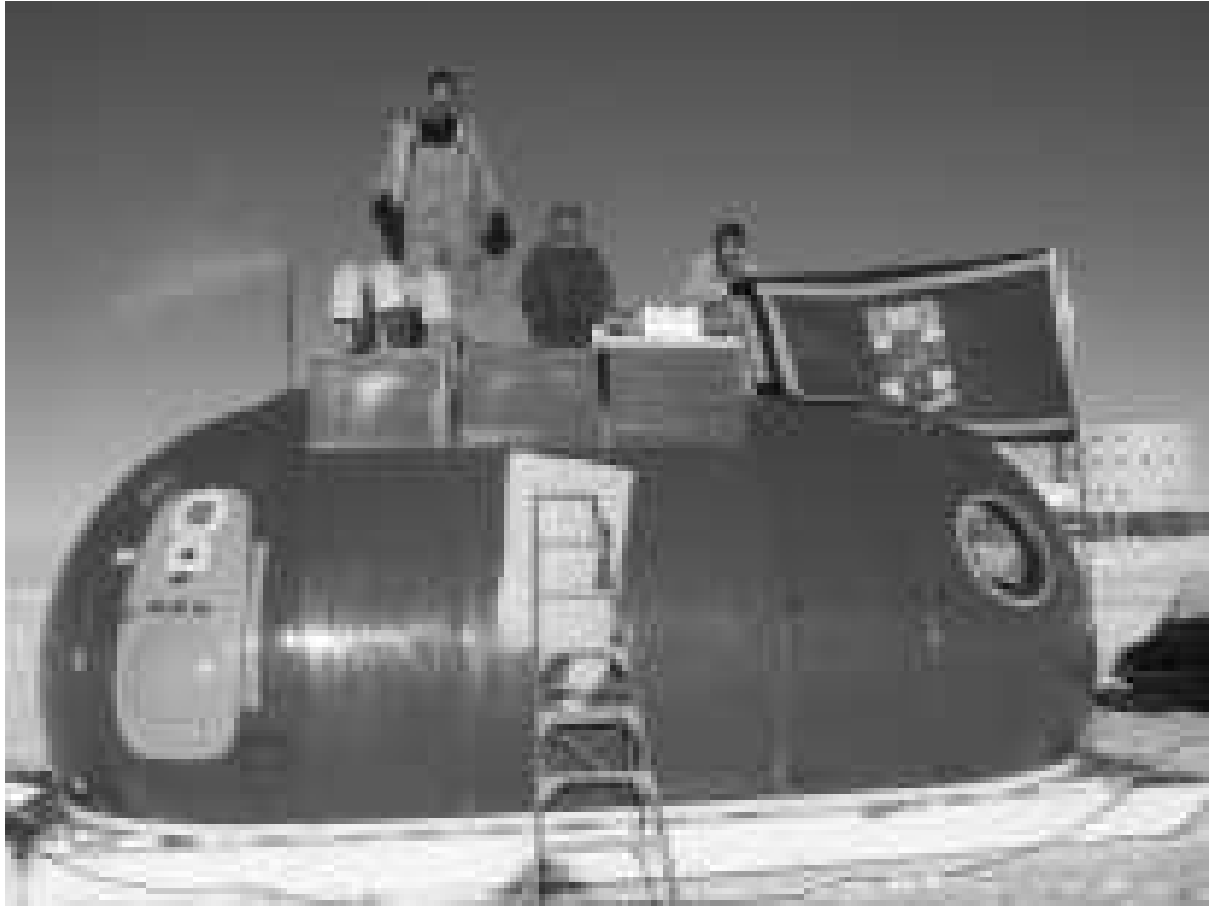,width=16cm}}
\caption{The AASTINO (Antarctic Astrophysical Site
Testing International Observatory), the successor to the AASTO.  The
AASTINO autonomously operated at Dome C through most of the winter of
2003, with the nearest human over 1,200\,km away. Contact was maintained
with its operators at the University of New South Wales in Sydney
through the Iridium satellite system. On top of the AASTINO are Jon
Lawrence, John Storey and Tony Travouillon, who installed the facility
in January 2003. {\scriptsize Image provided by John Storey, January
2003.}}
\label{AASTINO}
\end{figure}

The Dome C site at the French/Italian Concordia Station, due to open
for wintertime operation in 2005 (Candidi \& Ferrari 2003), is
currently being evaluated for its astronomy potential.  The AASTINO
(Lawrence et al.\ 2003; see Figure~\ref{AASTINO}) operated there
autonomously through most of the winter of 2003, over 1\,200\,km from
the nearest human, with communication made possible through the
Iridium satellite system. Dome C is one of the summits of the
Antarctic Plateau at an elevation of 3\,268\,m, and therefore the
katabatic winds, which are a constant feature at the South Pole (and
at all coastal locations), are minimal. In addition to the colder and
drier conditions than at the Pole, the atmospheric turbulence is
expected to be considerably smaller, thus leading to new opportunities
for adaptive optics correction of wavefronts, and for interferometry
(Lawrence 2004). 

The summit of the Plateau, the 4\,200\,m Dome A, is expected to
provide even better conditions for astronomy (Burton et al.\ 1994;
Storey et al.\ 2004), the best on the Earth for a wide range of
ground-based astronomy. There are no immediate plans by any nation to
construct a base there, and indeed no human has yet visited the site.
Nevertheless, plans have recently been advanced to establish an
automated facility there, such as the AASTINO\@.

\section{Conclusion}
\label{sec-conclusion}
This review of astrophysics at the bottom of the world has encompassed
almost a century of human endeavour, beginning with the discovery of
the Adelie Land Meteorite in 1912, during the `heroic' age of
Antarctic exploration, a period when humans first began to appreciate
the vastness and remoteness of the last continent.  It was to be
nearly fifty years before the next advances in astronomy were made there,
and it is little more than two decades 
since Antarctic
astrophysics began in earnest.  The 1990's have seen an avalanche of
activities, spanning the continent and in particular focussing on the
high inland plateau, where the conditions are particularly appealing
for a wide range of astronomical observations.  This review has paid
particular attention to the development of ``photon'' astronomy in
Antarctica, but equally ambitious have been some of the developments
in high energy astrophysics, seeking to measure the incident particle
fluxes from space.  The scope of this article does not allow for a
description of all the projects that have taken place, and a selection
of which to portray has had to be made by the authors.  A
comprehensive compilation of astronomical experiments that have been
conducted in Antarctica is, however, given in Table 1.  This provides
some further detail on experiments described here, as well as
information on a number of experiments not mentioned in the text.

Antarctica has already produced a range of stunning astronomical
results despite the relative infancy of most of the facilities that
have been operating.  This is particularly so for the cosmic microwave
background experiments, where the stable conditions have facilitated
producing results that compare very favourably to those later obtained
by satellites, as well as extending the range over which the
anisotropies can be probed.  There is no doubt that the summits of the
plateau provide superlative conditions for the conduct of a wide range
of observational astronomy.  It will be a fascinating story to follow
the development of new facilities on the high plateau in the third
millennium, and to wonder about the scientific problems they will be
able to tackle.


\section*{Acknowledgments}
Any attempt to produce an historical account of a subject as diverse
and expansive as Antarctic astronomy is not possible without the help
and advice of a great many people.  We particularly wish to thank
Paolo Calisse, Marc Duldig, Peter Gillingham, Stephen Haddelsey,
Per-Olof Hulth, Tony Stark and John Storey for their input and their
encouragement.  The authors have endeavoured to report the historical
facts of the development of the subject as best as they can obtain
them, but undoubtedly there will be some errors and omissions in this
work. We apologise for any inadvertent mistakes that we may have made,
and to anyone whose contribution has been omitted as a result.

\newpage
\section*{References}

\reference Ahn, H. S., et al. 2002, {\it Advanced Thin Ionization Calorimeter (ATIC) Update}, Am. Phys. Soc. Meeting (April 2002, Albuquerque), Abs B17.036

\reference Allen, C., et al. 1998, {\it Status of the Radio Ice Cherenkov Experiment (RICE)}, New Astronomy Reviews, 42, 319

\reference Allen, D. A., \& Crawford, J. W. 1984, {\it Cloud structure on the dark side of Venus}, Nature, 307, 222

\reference Allen, D. A., Hyland, A. R., \& Jones, T. J. 1983, {\it High-resolution images of the Galactic centre}, MNRAS, 204, 1145

\reference Andres, E., et al. 2000, {\it The AMANDA neutrino telescope: principle of operation and first results}, Astroparticle Phys., 13, 1

\reference Andres, E., et al. 2001, {\it Observation of high-energy neutrinos using Cerenkov detectors embedded deep in Antarctic ice}, Nature, 410, 441

\reference Andrewes, W. J. H. (Ed). 1996, The Quest for Longitude: Proceedings of the Longitude Symposium, 
(Harvard University)

\reference Aristidi, E., Agabi, A., Vernin, J., Azouit, M., Martin, F.,
Ziad, A., \& Fossat, E. 2003, {\it Antarctic site testing: first daytime seeing monitoring at Dome C}, A\&A, 406, L19

\reference Asakimori, K., et al. 1998,{\it Cosmic ray proton and helium spectra -- results from the JACEE}, ApJ, 502, 278

\reference Ashley, M. C. B. 1998,  {\it The near--IR and mid--IR 
sky background at the South Pole}, in ASP Conf. Ser. 141, 
`Astrophysics From Antarctica', eds. G. Novak \&
R. H. Landsberg (Astro. Soc. of the Pacific), 285

\reference Ashley, M. C. B., Burton, M. G., Storey, J. W. V., Lloyd, J. P., 
Bally, J., Briggs, J. W., \& Harper, D. A. 1996, {\it South Pole
observations of the near-infrared sky brightness}, PASP, 108, 721

\reference Ashley, M. C. B., Burton, M. G., Calisse, P. G., Phillips, A., \&
Storey, J. W. V. 2004, {\it Site testing at Dome C --- cloud statistics
from the ICECAM experiment}, in IAU Pub. Highlights in Astronomy, 13, 
eds. O. Engvold \& M. G. Burton (Astro. Soc. of the Pacific), in press


\reference Bally, J., Thiel, D., Billawala, Y., Potter, D., Loewenstein, 
R. F., Mrozek, F, \& Lloyd, J. P. 1996, {\it A Hartmann differential
image motion monitor (H--DIMM) for atmospheric turbulence
characterisation}, PASA, 13, 22

\reference Barwick, S. W., et al. 2003,  {\it Overview of the 
ANITA project}, SPIE, 4858, 265

\reference Bayly, P. G. W., \& Stillwell, F. L. 1923, The Adelie Land 
Meteorite, Scientific Reports Series A, Vol. IV, Geology (Sydney: A. J. Kent)

\reference Bayly, W., \& Cook, J. 1782, The original astronomical 
observations made in the course of a voyage to the Northern Pacific
Ocean, for the discovery of a North East or North West passage
(London: William Richardson)

\reference Bieber, J. W., Evenson, P., Dr\"{o}ge, W., Pyle, R., Ruffolo, D., Rujiwarodom, M., Tooprakai, P., \& Khumlumlert, T. 2004, {\it Spaceship Earth Observations of the Easter 2001 Solar Particle Event}, ApJ, 610, L103


\reference Boccas, M., Ashley, M. C. B., Phillips, M. A., Schinckel, 
A. E. T., \& Storey, J. W. V. 1998, {\it Antarctic Fiber Optic
Spectrometer}, PASA, 110, 306

\reference Boggs, S. E. et al. 1998,  {\it A high resolution 
gamma-ray and hard X-ray spectrometer (HIREGS) for long duration
balloon flights},  Adv. in Space. Res., 21, 1015

 
\reference Brooks, K. J., Burton, M. G., Rathborne, J. M., Ashley,
M. C. B., \& Storey, J. W. V. 2000, {\it Unlocking the Keyhole--H$_2$
and PAH emission from molecular clumps in the Keyhole Nebula}, MNRAS, 319, 95
 
\reference Burton, M. G. 1996, {\it Why Antarctica?}, PASA, 13, 2

\reference Burton, M. G., Allen, D. A., \& McGregor, P. J. 1993, {\it The 
potential of near--infrared astronomy in Antarctica},  
in the Proceedings of the Aust. Inst. Phys. 10th Congress, ANARE Research Notes 88,
eds. G. Burns \& M. Duldig (Aust. Ant. Div.), 293

\reference Burton, M. G., et al. 1994, {\it The scientific
potential for astronomy from the Antarctic plateau}, PASA, 11, 127

\reference Burton, M. G., et al. 1996, {\it JACARA's
plans}, PASA, 13, 33
 
\reference Burton, M. G., et al. 2000, {\it High-Resolution imaging of
photodissociation regions in NGC 6334}, ApJ, 542, 359

\reference Caldwell, D. A., Witteborn, F. C., Showen, R. L., Ninkov, Z., 
Martin, K. R., Doyle, L. R., \& Borucki, W. J. 2003, {\it The South
Pole Exoplanet Transit Search}, 25th IAU GA Special Session 2, `Astronomy in Antarctica'

\reference Calisse, P. G., Ashley, M. C. B., Burton, M. G., Phillips, M. A, 
Storey, J. W. V., Radford, S. J. E., \& Peterson, J. B. 2004, {\it
Submillimeter site testing at Dome C, Antarctica}, PASA, 21, 1


\reference Carlstrom, J. E. 2003, {\it The 8m South Pole Telescope}, 25th IAU GA Special Session 2, 
`Astronomy in Antarctica'

\reference Candidi, M. \& Ferrari, A. 2003, {\it Concordia Station at 
Dome C---a new site for astronomical research}, 25th IAU GA Special Session 2, `Astronomy in Antarctica'

\reference Chamberlain, M. A., Ashley, M. C. B., Burton, M. G., Phillips, A.,
\& Storey, J. W. V. 2000,  {\it Mid-infrared observing conditions at the
South Pole}, ApJ, 535, 501

\reference Chamberlin, R. A., Martin, B., Martin, C., \& Stark, A. A.
2003, {\it A sub-millimeter atmospheric FTS at the geographic South
Pole}, in SPIE Proc. 4855, `MM and sub--mm detectors for astronomy', 
eds. T.G. Phillips \& J. Zmuidzinas, 609

\reference Chen, K.--Y., Oliver, J. P., \& Wood F. B. 1987, {\it Stellar
Photometry at the South Pole Optical Telescope}, Antarctic J. of the US,
1987 Review Vol. XXII, No. 5, 283

\reference Coble, K., et al.  1999, {\it Anisotropy in the cosmic microwave background at degree
angular scales: Python V results}, ApJ, 519, L5

\reference Cordaro, E. G., \& Storini, M. 1992,  {\it Cosmic-ray measurements 
in Antarctica during the international Solar-Terrestrial Energy
Program}, Nuovo Cimento C, Serie 1, 
15, 539

\reference dall'Oglio, G., \& de Bernardis, P. 1988, {\it Observations of 
cosmic background radiation anisotropy from Antarctica}, ApJ, 331, 547

\reference de Bernardis, P., et al. 2000,  {\it A flat Universe 
from high-resolution maps of the cosmic microwave background
radiation}, Nature, 404, 955

\reference Dempsey, J. T., Storey, J. W. V., \& Ashley, M. C. B. 2003a, 
{\it COBBER: a pocket cloud detector for Dome C}, 
Memorie della Societa Astronomica Italiana Supplement, 2, 70

\reference Dempsey, J. T., Storey, J. W., Ashley, M. C., Burton, M. G., 
Jarnyk, M. \& Hovey, G., 2003b, {\it AFOS: probing the UV-visible
potential of the Antarctic plateau}, 25th IAU GA Special Session 2, 
`Astronomy in Antarctica'

\reference Dickinson, J. E., et al. 2000,  {\it The New South Pole Air Shower
Experiment -- SPASE--2}, Nucl. Instr. and Meth. A, 440, 95

\reference Dopita, M. A., Wood, P. R., \& Hovey, G. R. 1996, {\it 
An automated DIMM telescope for Antarctica}, PASA, 13, 39

\reference Dragovan, M., Stark, A. A., Pernic, R., \& Pomerantz, M.
1990, {\it Millimetric sky opacity measurements from the South Pole}, J. Appl. Optics, 29, 463

\reference Duldig, M. 2002, {\it Cosmic ray physics and astronomy}, 
in `Australian Antarctic Science: the 
first 50 years of ANARE', eds. H.J. Marchant, D.J. Lugg \& P.G. Quilty 
(Aust. Ant. Div.), 43


\reference Engargiola, G., Zmuidzinas, J., \& Lo, K.--Y. 1994, {\it A
492 GHz quasioptical SIS receiver for submillimeter astronomy}, 
Rev. Sci. Instr., 65, 1833

\reference Fossat, E., Gelly, B., Grec, G., \& Pomerantz, M. 1987,  {\it
Search for Solar P-Mode frequency changes between 1980 and 1985}, A\&A, 177, L47

\reference Fowler, A. M., et al. 1998, {\it Abu/SPIREX: the South Pole thermal
IR experiment}, SPIE Proc., 3354, 1170


\reference Gillingham, P. R. 1991, {\it Prospects for an Antarctic 
observatory}, PASA, 9, 55

\reference Gillingham, P. R. 1992, Report of Joint Commission Meeting No.\ 4, 
IAU Highlights of Astronomy, 9, ed. J. Bergeron (Kluwer Acad. Pub.), 575

\reference Gillingham, P. R. 1993a, {\it Super-seeing from the Australian
Antarctic Territory}, in the Proceedings of the Aust. Inst. Phys. 
10th Congress, ANARE Research Notes 88, eds. G. Burns \& M. Duldig (Aust. Ant. Div.), 290

\reference Gillingham, P. R. 1993b,  {\it Super-seeing from Antarctica}, 
in Proc. 32nd Herstmonceux Conf,
`Optics in Astronomy', ed. J.V. Wall (Cambridge University Press), 244

\reference Grec, G., Fossat, E., \& Pomerantz, M. 1980, {\it Solar 
oscillations: full disk observations from the geographic South Pole}, 
Nature, 288, 541

\reference Grec, G., Fossat, E., \& Pomerantz, M. 1983, {\it Full disk
observations of Solar oscillations from the geographic South Pole:
latest results}, Solar Physics, 82, 55

\reference Gurney, A. 1998, Below the Convergence Voyages Toward Antarctica 1699-1839
(Penguin)

\reference Halzen, F. 2004, {\it IceCube: a kilometer-scale neutrino
observatory at the South Pole}, in IAU Pub. Highlights in Astronomy, 13, 
eds. O. Engvold \& M. G. Burton (Astro. Soc. of the Pacific), in press

\reference Harper, D. 1989, {\it Infrared astronomy in Antarctica}, 
in AIP Conf. Proc., 198, `Astrophysics in Antarctica', 
ed. D.J. Mullan, M.A. Pomerantz \& T. Stanev (American Inst. Phys.), 123

\reference Harper, D. A. 1994, {\it Infrared arrays on ice: new opportunities 
for astronomy in Antarctica}, in Astrophys. \& Sp. Sci. Lib., 190,
`Infrared Astronomy with Arrays, the Next Generation', ed. 
I. McLean (Kluwer Acad. Pub.), 247

\reference Harvey, J. 1989, {\it Solar observing conditions at the South 
Pole}, in AIP Conf. Proc., 198, `Astrophysics in Antarctica', 
ed. D.J. Mullan, M.A. Pomerantz \& T. Stanev (American Inst. Phy.), 227

\reference Hereld. M. 1994, {\it SPIREX -- near--IR astronomy from the
South Pole}, in Astrophys. \& Sp. Sci. Lib., 190,
`Infrared Astronomy with Arrays, the Next Generation', ed. 
I. McLean (Kluwer Acad. Pub.), 248

\reference Ingalls, J. G., Bania, T. M., Lane, A. P., Rumitz, M., \&
Stark, A. A. 2000, {\it Physical state of molecular gas in high
galactic latitude translucent clouds}, ApJ, 535, 211
 
\reference Jacklyn, R. M. 2000, {\it Underground studies in Tasmania and at
Mawson}, in ANARE Research Notes 102, 
`50 Years of Cosmic Ray Research in Tasmania and Antarctica', 
ed. M. Duldig (Aust. Ant. Div.), 91

\reference Jacklyn, R. M. \& Cooke, D. J. 1971,  {\it Components of the sidereal
anisotropy}, in Proc. 12th Int. Cosmic Ray Conf., vol 1, 290

\reference Jacklyn, R. M., Duldig, M. L., \& Pomerantz, M. A. 1987, {\it
High-energy cosmic ray intensity waves}, JGR, 92, 8511




\reference Keating, B. G. et al. 2003, {\it BICEP--a
large scale CMB polarimeter}, in SPIE Proc., 4843, 
`Polarimetry in Astronomy', 284

\reference Kenyon, S. J., \& Gomez, M. 2001,  {\it A 3 micron survey of the 
Chamaeleon I dark cloud}, AJ, 121, 2673
 
\reference Kovac, J. M., Leitch, E. M., Pryke, C., Carlstrom, J. E., 
Halverson, N. W., \& Holzapfel, W. L. 2002,  {\it Detection of
polarization in the cosmic microwave background using DASI}, Nature, 420, 772

\reference Kowitt, M. S., et al. 1995, {\it The MSAM/TopHat program of anisotropy measurements},
Astrophys. Lett. Comm., 32, 273

\reference Kuo, C. L. et al. 2003, {\it High resolution 
observations of the cosmic microwave power spectrum with ACBAR}, ApJ, 600, 32


\reference Lane, A. P. \& Stark, A. A. 1997, {\it Antarctic 
Submillimeter Telescope and Remote Observatory (AST/RO): installation
at Pole}, Antarctic J. of the US, 30, 377

\reference Law, P. 2000, {\it Cosmic rays in the Antarctic--laying the
foundations},  in ANARE Research Notes 102, 
`50 Years of Cosmic Ray Research in Tasmania and Antarctica', 
ed. M. Duldig (Aust. Ant. Div.), 27

\reference Lawrence, J. S. 2004,  {\it Infrared and sub-millimetre atmospheric
characteristics of high Antarctic plateau sites}, PASP, in press

\reference Lawrence, J. S., et al. 2002, {\it Operation of the Near Infrared Sky Monitor at the South
Pole}, PASA, 19, 328

\reference Lawrence, J. S., et al. 2003,{\it The AASTINO: Automated Astrophysical Site
Testing International Observatory},  Memorie della Societa Astronomica Italiana Supplement, 2, 217

\reference Leitch, E. M., et al. 2002, {\it Measurement of polarization with the
Degree Angular Scale Interferometer}, Nature, 420, 763

\reference Link, J. T. et al. 2002, {\it Preliminary results
from the 2001-2002 balloon flight of the TIGER cosmic ray detector}, Am. Phys. Soc. Meeting 
(April 2002, Albuquerque), Abs P11.002

\reference Lloyd, J. P., Oppenheimer, B. R., \& Graham, J. R. 2002,  {\it
The potential of differential astrometric interferometry from the high
Antarctic plateau}, PASA, 19, 318

\reference Loewenstein, R. F., Bero, C., Lloyd, J. P., Mrozek, F.,
Bally, J., \& Theil, D. 1998, {\it Astronomical seeing at the South
Pole}, in ASP Conf. Ser. 141, `Astrophysics From Antarctica', eds. G. Novak \&
R. H. Landsberg (Astro. Soc. of the Pacific), 296

\reference Lyo, A.-R., Lawson, W. A., Mamajek, E. E., Feigelson, E. D.,
Sung, E.-O., \& Crause, L. A. 2003,  {\it Infrared study of the Eta
Chamaeleontis cluster and the longetivity of circumstellar disks}, MNRAS, 338, 616

\reference Marks, R. D. 2002, {\it Astronomical seeing from the summits of 
the Antarctic plateau}, A\&A, 385, 328

\reference Marks, R. D., Vernin, J., Azouit, M., Briggs, J. W., Burton, M. G., 
Ashley, M. C. B., \& Manigault, J. F. 1996, {\it Antarctic site
testing -- microthermal measurements of surface-layer seeing at the
South Pole}, A\&AS, 118, 385

\reference Marks, R. D., Vernin, J., Azouit, M., Manigault, J. F., \& 
Clevelin, C. 1999, {\it Measurements of optical seeing on the high
Antarctic plateau}, A\&AS, 134, 161

\reference Martin, C. L., Walsh, W. M., Xiao, K., Lane, A. P., Walker, C. K., 
\& Stark, A. A. 2004,  {\it The AST/RO survey of the Galactic center
region. I: the inner 3 degrees},  ApJ, submitted

\reference Mawson, D. 1915, The home of the blizzard: being the story of 
the Australasian Antarctic Expedition, 1911-1914 (London: William Heinemann)

\reference McCracken, K. G.  1962,  {\it The cosmic-ray flare effect: 3, 
deductions regarding the interplanetary magnetic field}, JGR, 67, 447

\reference McCracken, K. G. 2000,  {\it The Australian Neutron Monitor 
Network}, in ANARE Research Notes 102, 
`50 Years of Cosmic Ray Research in Tasmania and Antarctica', 
ed. M. Duldig (Aust. Ant. Div.) 81

\reference Morse, R., \& Gaidos, J. 1989,  {\it A South Pole facility to 
observe very high energy gamma ray sources}, in AIP Conf. Proc., 198, 
`Astrophysics in Antarctica', ed. D.J. Mullan, M.A. Pomerantz \& T. Stanev 
(American Inst. Phys.), 24

\reference Mosconi, M., Recabarren, P., Ferreiro, D., Renzi, V., \& Ozu, R.
1990, {\it Reporte de actividades de la Estaci\'{o}n Astron\'{o}mica Polar
``J.L.Sersi''}, Boletin de la Asociacion Argentina de Astronomia, 42, 67

\reference Mullan, D. J., Pomerantz, M. A. \& Stanev, T. (Eds).
1989, AIP Conf. Proc., 198, `Astrophysics in Antarctica' (American Inst. Phys.)

\reference Nagata, T. (Ed). 1975, {\it Yamato meteorites collected in 
Antarctica in 1969}, Memoirs of National Institute of Polar
Research, Special Issue No.\ 5 


\reference Nguyen, H. T., et al. 1996,  {\it
The South Pole near infrared sky brightness}, PASP, 108, 718

\reference Novak, G., et al. 2003, {\it First results from the Submillimeter Polarimeter for
Antarctic Remote Observations: evidence of large-scale toroidal
magnetic fields in the Galactic center}, ApJ, 583, L83


\reference Ojha, R., et al. 2001, {\it AST/RO observations of atomic carbon near the
Galactic center}, ApJ, 548, 253-257

\reference Opdishaw, H. 1957, in Proceedings of the 
symposium at the opening of the International Geophysical Year, 
ed. H. Opdishaw et al., Geophysical monograph series, 2

\reference Pajot, F., et al. 1989,  {\it Observations of the submillimetre integrated galactic
emission from the South Pole}, A\&A, 223, 107

\reference Parsons, N. R. 1957, {\it The design and operation of ANARE cosmic ray detector
``C''}, ANARE Internal Reports, 17. Pub. \#36 (Aust. Ant. Div.), 57

\reference Parsons, N. R. 2000,  {\it Cosmic Ray Observations at Mawson -- 
the early days}, in ANARE Research Notes 102, `50 Years of Cosmic Ray
Research in Tasmania', ed. M. Duldig (Aust. Ant. Div.), 57

\reference Peary, R. E. \& Frost, E. B. 1912, private communication with 
J. W. Briggs and J. Bausch at Yerkes Observatory, Spring 2002, about a
letter exchange between Peary and Frost in 1912

\reference Peterson, J. B., et al. 2000, {\it First results from Viper: detection
of small-scale anisotropy at 40 GHz}, ApJ, 532, L83

\reference Phillips, A., Burton, M. G., Ashley, M. C. B., Storey, J. W. V., 
Lloyd, J. P., Harper, D. A., \& Bally, J. 1999, {\it The
near-infrared sky emission at the South Pole in winter}, ApJ 527, 1009

\reference Piccirillo, L., et al. 2002, {\it QUEST--a 2.6\,m mm-wave
telescope for CMB polarimetric studies}, in AIP Conf. Proc., 609,
`Astrophysical Polarized Backgrounds', ed. S. Cecchini, S. Cortiglioni, R. Sault 
\& C. Sbarra (American Inst. Phys), 159

\reference Pomerantz, M. A. 1986, {\it Astronomy on Ice}, Proc. A.S.A., 6, 403

\reference Pomerantz, M. A. 2000,  {\it Interview with Dr.\ M. A. 
Pomerantz conducted in his San Rafael, California home on May 10, 2000}, 
in the Polar Oral History Program, 
Ohio State University Archives. Oral history interview with Dr.\ Martin Pomerantz, 
conducted by Brian Shoemaker, May 10, 2000

\reference Pomerantz, M. A., Wyller, A. A., \& Kusoffsky, U. 1981, {\it The polar solar observatory}, 
Antarctic J. of the US, 16, 221

\reference Pomerantz, M. A., Harvey, J. W., \& Duvall, T. 1982, {\it Large scale motions
and the structure of the Sun}, Antarctic J. of the US, 17, 232

\reference Rathborne, J. M., Burton, M. G., Brooks, K. J., Cohen, M., Ashley, 
M. C. B., \& Storey, J. W. V. 2002, {\it Photodissociation regions and
star formation in the Carina nebula}, MNRAS, 331, 85

\reference Rathborne, J. M., \& Burton, M. G. 2004, {\it 
Results from the South Pole InfraRed EXplorer Telescope}, in IAU Pub. Highlights in Astronomy, 13, 
eds. O. Engvold \& M. G. Burton (Astro. Soc. of the Pacific), in press

\reference Ravick, M. G., \& Revnov, B. I. 1965, {\it Lazarev iron meteorite}, Meteoritika, 23, 38

\reference Runyan, M. C., et al. 2003,  {\it The Arcminute 
Cosmology Bolometer Array},  ApJS, 149, 265

\reference Rust, D. M. 1994,  {\it The Flare Genesis Project}, Adv. Space Res., 14, 89

\reference Severson, S. 2000, `Death of a comet: SPIREX observations of 
the collision of SL9 with Jupiter', PhD Dissertation, Univ.\ of
Chicago

\reference Shea, M. A., \& Smith, D. F. 2000, {\it Fifty years of cosmic 
radiation data}, Space Science Reviews, 93, 229

\reference Shima M. \& Shima M. 1973, {\it Mineralogical and chemical composition of new
Antarctica meteorites}, Meteoritics, 8, 439

\reference Smith, C. H., \& Harper, D. A. 1998, {\it Mid-infrared 
sky brightness site testing at the South Pole}, PASP, 110, 747

\reference Smith, N. J. T., et al. 1989, {\it An experiment to search for ultra high energy gamma
ray sources from the South Pole},  Nucl. Inst. and Meth. A, 276, 622

\reference Spergel, D. N., et al. 2003, {\it First-Year Wilkinson Microwave 
Anisotropy Probe (WMAP) observations: determination of cosmological
parameters}, ApJS, 148, 175

\reference Stark, A. A., Bolatto, A. D., Chamberlin, R. A., Lane, 
A. P., Bania, T. M., Jackson, J. M., \& Lo, K.--Y. 1997, {\it First
detection of 492 GHz [CI] emission from the Large Magellanic Cloud}, ApJ, 480, L59

\reference Stark, A. A., et al. 2001, {\it The Antarctic 
Submillimeter Telescope and Remote Observatory (AST/RO)}, PASA, 113, 567

\reference Storey, J. W. V. 2000,  {\it Infrared astronomy: in the heat of 
the night}, PASA, 17, 270

\reference Storey, J. W. V., \& Hyland, A. R. 1993, {\it Far-infrared and 
sub-millimeter astronomy in Antarctic}, in the Proceedings of 
the Aust. Inst. Phys. 10th Congress, ANARE Research Notes 88,
eds. G. Burns \& M. Duldig (Aust. Ant. Div.), 309

\reference Storey, J. W. V., Ashley, M. C. B., \& Burton, M. G. 1996, 
{\it An automated astrophysical observatory for Antarctica}, PASA, 13, 35

\reference Storey, J. W. V., Ashley, M. C. B., Boccas, M., Phillips, M. A., \& 
Schinckel, A. E. T. 1999, {\it Infrared sky brightness monitors for
Antarctica}, PASP, 111, 765

\reference Storey, J. W. V., Ashley, M. C. B., \& Burton, M. G. 2000, {\it 
Novel instruments for site characterization}, in
Proc. SPIE, 4008, `Astronomical telescopes and instrumentation 2000: 
optical and infrared telescope instrumentation and detectors', 1376

\reference Storey, J. W. V., Ashley, M. C. B., Burton, M. G., \& Lawrence, J.
2004, {\it Beyond Dome C}, in IAU Pub. Highlights in Astronomy, 13, 
eds. O. Engvold \& M. G. Burton (Astro. Soc. of the Pacific), in press

\reference Swain, M. R., Bradford, C. M., Stacey, G. J., Bolatto, A. D., 
Jackson, J. M., Savage, M. L., \& Davidson, J. A. 1998, {\it Design of
the South Pole Imaging Fabry-Perot interferometer (SPIFI)}, SPIE, 3354, 480

\reference Taylor, M. J. 1990,  {\it Photometry of the 4686\AA\ emission 
line of Gamma 2 Velorum from the South Pole}, AJ, 100, 1264

\reference Tolstikov, E. 1961,  {\it Discovery of Lazarev iron meteorite, 
Antarctica}, The Meteoritical Bulletin, 20, 1

\reference Tosti, G., Busso, M., Ciprini, S., Persi, P., 
Ferrari-Toniolo, M., \& Corcione, L. 2003,  {\it IRAIT: a telescope for
infrared astronomy from Antarctica}, Memorie della Societa Astronomica Italiana, 74, 37

\reference Townes, C. H., \& Melnick, G. 1990,  {\it Atmospheric 
transmission in the far-infrared at the South Pole and astronomical
applications}, PASP, 102, 357

\reference Travouillon, T., Ashley, M. C. B., Burton, M. G., Storey, 
J. W. V., \& Loewenstein, R. F. 2003a, {\it Atmospheric turbulence at
the South Pole and its implications for astronomy}, A\&A, 400, 1163

\reference Travouillon, T., et al. 2003b, {\it Automated Shack-Hartman seeing 
measurements at the South Pole},  A\&A, 409, 1169

\reference Travouillon, T., Ashley, M. C. B., Burton, M. G., Lawrence, J., 
\& Storey, J. W. V. 2003c, {\it Low atmosphere turbulence at Dome C: 
preliminary results},  Memorie della Societa Astronomica Italiana Supplement, 2, 150

\reference Tucker, G. S., Griffin, G. S., Nguyen, H. T., \& Peterson, J. B.
1993, {\it A search for small-scale anisotropy in the cosmic
microwave background}, ApJ, 419, L45

\reference Valenziano, L., et al. 1998, {\it APACHE96 CMBR anisotropy
experiment at Dome C},  in ASP Conf. Ser. 141, 
`Astrophysics From Antarctica', eds. G. Novak \&
R. H. Landsberg (Astro. Soc. of the Pacific), 81

\reference van Stekelenborg, J., et al. 1993, {\it
Search for point sources of ultrahigh energy gamma rays in the
southern hemisphere with the South Pole air shower array}, Phys. Rev. D, 48, 4504

\reference von Walden, P., \& Storey, J. W. V. 2003,  {\it First measurements of
the infrared sky brightness at Dome Concordia, Antarctica},
Memorie della Societa Astronomica Italiana, 2

\reference Vrana, A., \& Riley, J. F. 1968,  {\it Cosmic ray records, Mawson 1968},
ANARE Data Reports Series C(2), 111, (Aust. Ant. Div.),  109

\reference Walker, C. K., Kooi, J. W., Chan, M., LeDuc, H. G., Schaffer, P. L.,
Carlstrom, J. E., \& Phillips, T. G. 1992,  {\it A low-noise 492 GHz SIS
waveguide receiver}, Int. J. IR \& MM Waves, 13, 785

\reference Walker, C. K. et al. 2001, {\it POLE STAR: An 810 GHz Array Receiver for AST/RO}, 
in Proceedings of the 12th International Symposium on Space THz Technology, ed. I. Mehdi et al.

\reference Westphal, J. 1974, Final Report, NASA NGR 05-002-185

\reference Wiebusch, C. H., et al. 2002,  {\it Results from AMANDA}, Mod. Phys. Lett. A., 17, 2019

\reference Yan, M., Lane, A. P., \& Stark, A. A. 2004, {\it AST/RO Mapping of 
NGC 6334 in the CI 492 GHz and CO (J=4--3) 460 GHz Line}, in preparation

\reference Yngvesson, K. S., et al.  2001, {\it Terrahertz
receiver with NbN HEB device (TREND) -- a low noise receiver user instrument
for AST/RO at the South Pole}, in Proceedings of the 12th International 
Symposium on Space THz Technology, ed. I. Mehdi et al.

\reference Yoshida, M., Ando, H., Omoto, K., Naruse, R., \& Ageta, Y. 1971, 
{\it Discovery of meteorites near Yamato Mountains, East Antarctica}, Japanese Antarctic Record, 39, 62

\reference Zhang, X., Lee, Y., Bolatto, A. P., \& Stark, A. A.
2001, {\it CO (J=4--3) and [CI] observations of the Carina molecular
cloud complex}, ApJ, 553, 274



\newpage
\section{Tables and Figures}

\end{document}